\documentclass[epj]{svjour}
\usepackage{graphicx}
\input{epsf}
\usepackage{amssymb}
\usepackage{url}
\usepackage{color}
\usepackage{t1enc}
\usepackage{graphicx}
\usepackage{color}
\usepackage{amsmath}
\usepackage{amsfonts}
\usepackage{enumerate}
\usepackage{longtable}
\usepackage{dcolumn}
\usepackage{bm}
\usepackage{epsfig}
\usepackage{hyperref}
\usepackage{caption}
\usepackage{subcaption}
\usepackage{wrapfig}
\usepackage{bbm}

\def\rmse{\hbox{\textsc{rmse}}{}}
\def\err{\mbox{err}} 

\begin{document}

\title{Statistical analysis of NOMAO customer votes for spots of France}

\author{
R\'obert P\'alovics\inst{1,2}
\and
B\'alint Dar\'oczy\inst{1,2}
\and
Andr\'as Bencz\'ur\inst{1,3}
\and
Julia Pap\inst{1}
\and
Leonardo Ermann\inst{4}
\and
Samuel Phan \inst{5}
\and
Alexei D. Chepelianskii \inst{6}
\and
Dima L. Shepelyansky\inst{7}
}
\institute{
Informatics Laboratory, Institute for Computer Science and Control,\\
Hungarian Academy of Sciences (MTA SZTAKI), Pf. 63, H-1518 Budapest, Hungary
\and
Technical University Budapest, Hungary
\and
E\"otv\"os University Budapest, Hungary
\and
Departamento de F\'isica Te\'orica, GIyA, CNEA, Av. Libertador 8250, 
(C1429BNP) Buenos Aires, Argentina.
\and
NOMAO.COM, 1 av Jean Rieux, 31500 Toulouse, France 
\and
LPS, Universit\'e Paris-Sud, CNRS, UMR 8502, F-91405, Orsay, France
\and
Laboratoire de Physique Th\'eorique du CNRS, IRSAMC, 
Universit\'e de Toulouse, UPS, F-31062 Toulouse, France
}

\titlerunning{The spectrum of a geo-located rating matrix}

\abstract{We investigate the statistical properties of votes of customers
for spots of France collected by the startup company NOMAO.
The frequencies of votes per spot and per customer
are characterized by a power law distributions
which remain stable on a time scale of a decade when the number of votes
is varied by almost two orders of magnitude. Using the 
computer science methods 
we explore the spectrum and the eigenvalues of a matrix containing 
user ratings to geolocalized items. 
Eigenvalues nicely map to large towns and regions but show certain 
level of instability as we modify the interpretation of the underlying matrix.
We evaluate imputation strategies that provide improved prediction performance 
by reaching geographically smooth eigenvectors.
We point on possible links 
between distribution of votes and the 
phenomenon of self-organized criticality.
}


\date{Dated:  May 5, 2015}

\maketitle

\section{Introduction}

\begin{figure}
          \centering
          \includegraphics[width=\columnwidth]{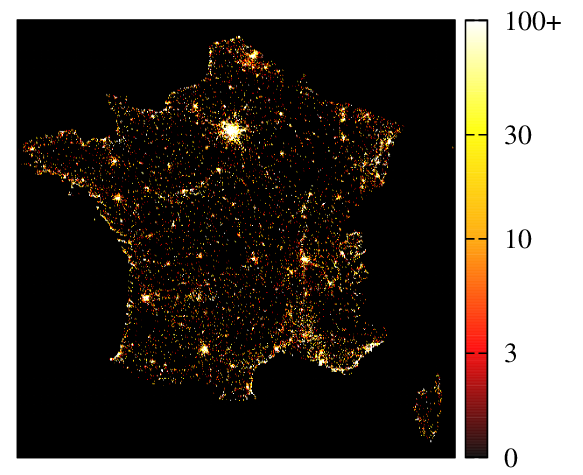}
          \includegraphics[width=\columnwidth]{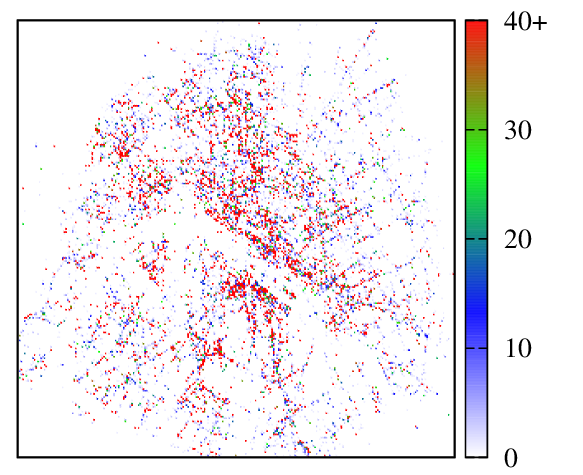}
\caption{Geographical distribution of votes for spots (POIs) in the original datasets. 
Top panel: case of France; 
(each square pixel represents  $7.8 {km}^2$); bottom panel: case of Paris
(each square pixel represents  $1370  {m}^2$); 
color bars give a number of votes per pixel (cell), 
a limitation  in number of votes 
is introduced for a better color representation.} 
\label{fig:votes}
\end{figure}

\begin{figure}
\begin{center}
\includegraphics[width=0.48\textwidth]{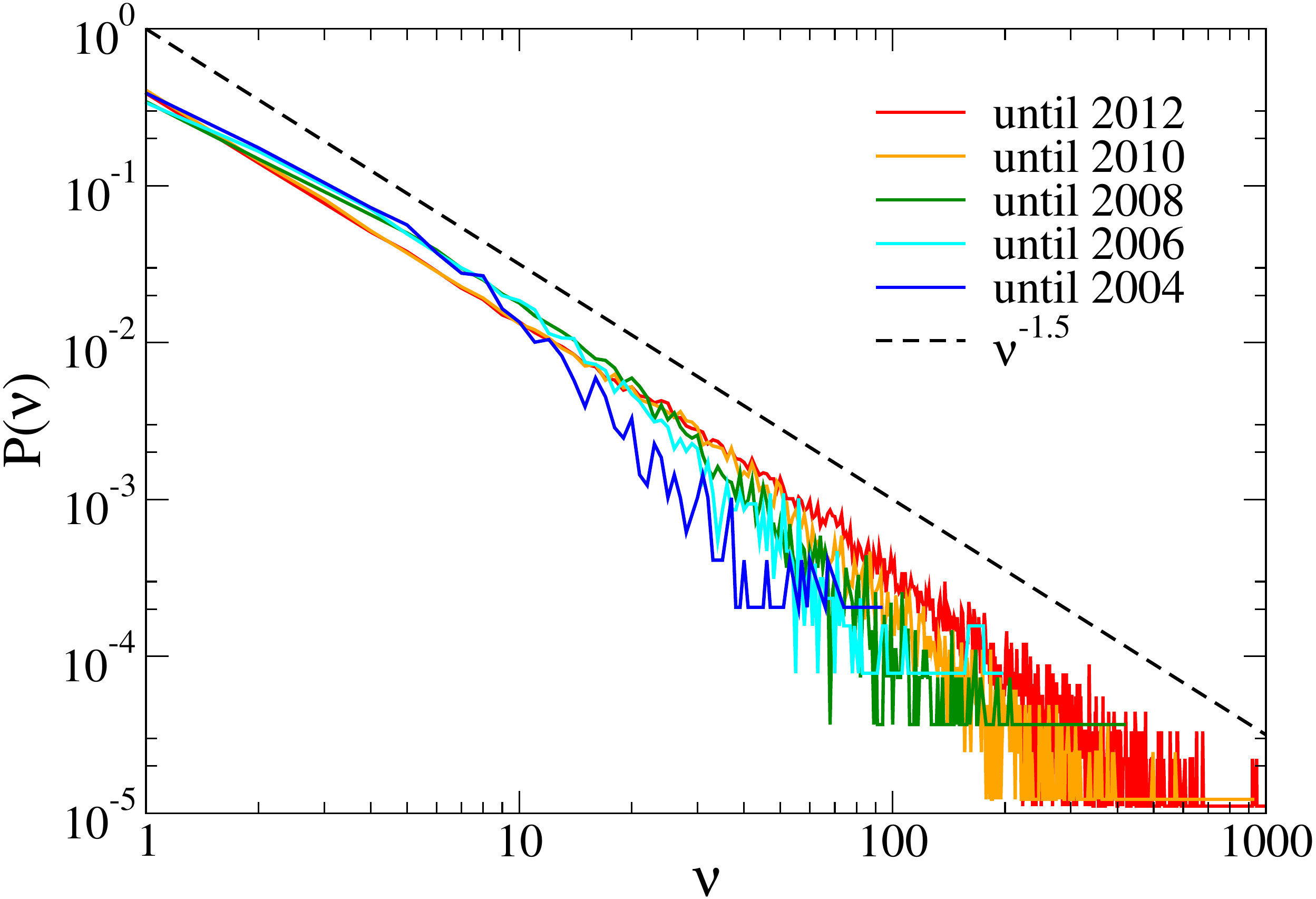}\\
\includegraphics[width=0.48\textwidth]{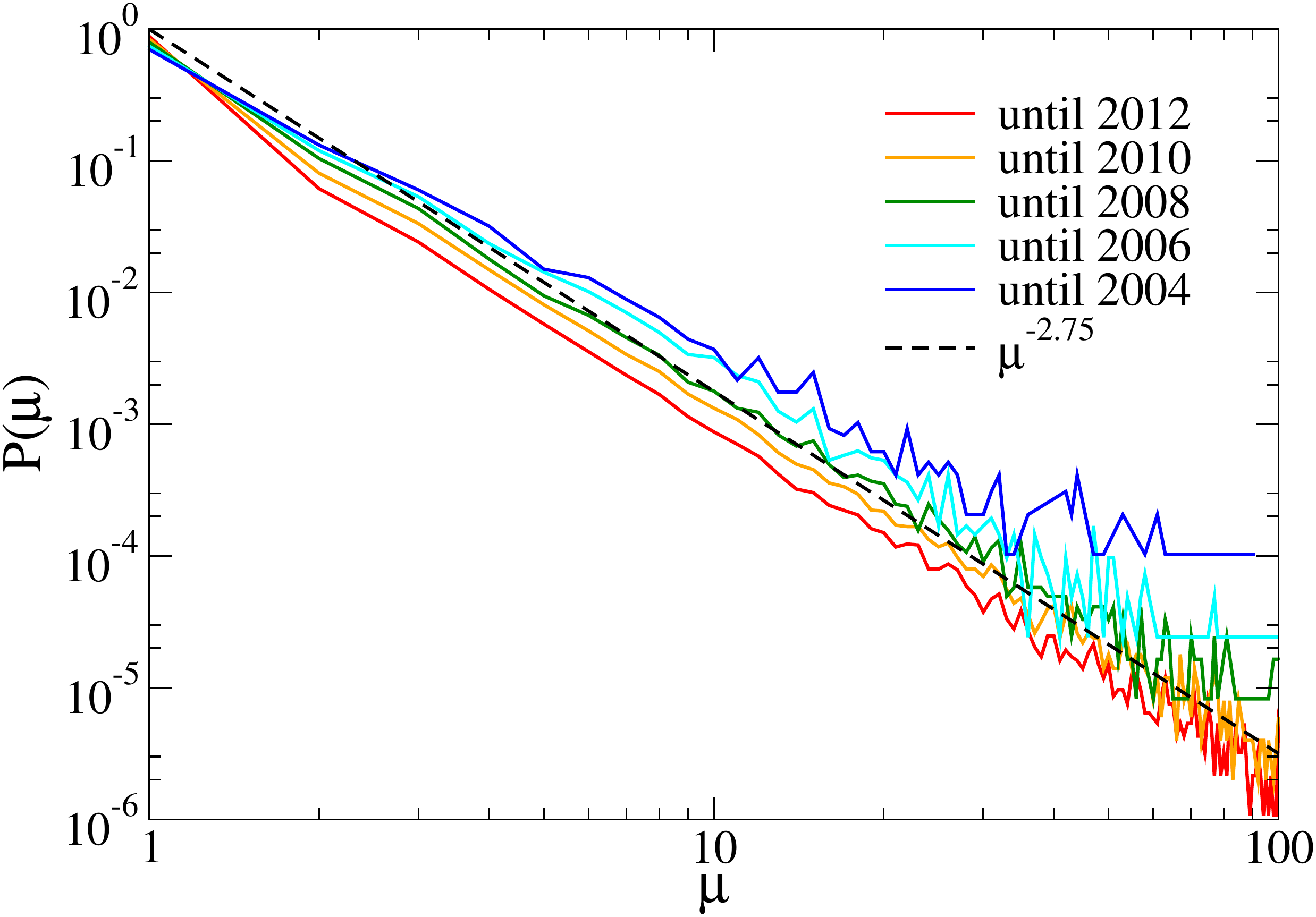}
\end{center}
\vglue -0.1cm
\caption{Differential frequency distributions of votes for the case of France
for different time intervals until year 2004, 2006, 2008, 2010, 2012.
Top panel: differential probability $P(\nu)$ to have 
$\nu$ votes for spots (POIs); bottom panel:  differential probability $P(\mu)$ to have 
$\mu$ votes per user (customer). Here the dashed lines show an average 
algebraic decay with exponents $a=1.5$ (top), $b=2.75$ (bottom).} 
\label{figtimevol}
\end{figure}

The young startup company NOMAO \cite{nomaocom} collected a large database about customer
(or user) votes for spots (or Points of Interest POIs or items) in France.
The spots represent mainly restaurants and hotels
with known geolocation coordinates.
In this paper we investigate 
the statistical properties of these NOMAO votes and
ratings of geolocalized items  
in a mix of geographic information and recommendations systems.  
The geographical distributions of votes are shown in Fig.~\ref{fig:votes}
for the whole France and more specifically for Paris.
The frequency distributions of votes per spot and votes per user
are shown in Fig.~\ref{figtimevol}
for France at different time intervals. It shows that these
frequency distributions are stabilized in time
and thus we are dealing with an unusual statistical system
been at a certain  steady-state. We note that at present a 
variety of real systems and networks are found to possess
power law distribution (see e.g. \cite{dorogovtsev})
and thus here we investigate a new type of such a case
with algebraic statistical properties.

To analyze the statistical properties of this real system
we use the methods of recommender systems \cite{wikirecommender}
which gained a broad recognition in computer science
after the Netflix Prize competition \cite{wikinetflix}.
In our research, distance, region and location become a side information over 
a multi-objective classification or regression problem. We concentrate on predicting 
user preferences by a spectral analysis based collaborative filtering that 
uses geo-location in addition to the ratings matrix.

We investigate how user taste, as described by latent factors, 
is reflected in the geographic information system.
We compare the latent factors obtained by a full spectral analysis 
and by the stochastic gradient method, the standard recommendation 
technique \cite{wikirecommender}
applicable for matrices with a very large fraction of missed values. 


The key difficulty in the spectral analysis lies in the abundance of 
missing values in the rating matrix: our matrix consists of 99.5\% missing values 
while the Netflix matrix for example is 99\% unknown. Several early results describe 
expectation maximization based singular value decomposition (SVD) algorithms,
dating back to the seventies \cite{gabriel79weights} and 
\cite{canny02collaborative,srebro03weighted,zhang05svd}, 
describe the method for a recommender application.  

A successful implementation of spectral analysis in recommender matrices 
with only a few known elements is described by Simon
Funk in \cite{funk06netflix}.  His method is a variant of 
Stochastic Gradient Descent (SGD) reminiscent of gradient boosting \cite{friedman99gradient}.   
SGD computes no eigenvalues and does not guarantee the orthogonality of the matrix factors.  
On the other hand, regularization is easily incorporated in SGD, which enables 
a better handling of the very large amount of missing values in the matrix and 
in particular, prevents overfit to training elements and provide better 
quality predictions of the unknown ratings.

In this paper, after describing methods and related results (Section~2) and 
the NOMAO data sets (Section~3), 
we compare and visualize the geo-localization of the matrix factors 
defined by SVD and SGD under various parameter settings in 
Section~\ref{sect:reco-matrix-spectrum}. We show that by imputing ratings 
to nearby locations we may form factors that yield a better description 
of the ratings matrix in Section~\ref{sec:collaborative_filtering}.

\section{Methods and related results}

Recommenders based on the rank $k$ approximation of the rating matrix with
the first $k$ singular vectors are
probably first described in 
\cite{billsus98collaborative,pryor98svd,gupta99jester,sarwar00application} 
and many others near year 2000.

The Singular Value Decomposition (SVD) of a rank $\rho$ matrix $R$ is given 
by $R=U^T \Sigma V$ with $U$ an ${m\times\rho}$, $\Sigma$ a ${\rho\times\rho}$ 
and $V$ an ${n\times\rho}$ matrix such that $U$ and $V$ are orthogonal.
By the Eckart-Young theorem  \cite{GolubVanLoan} the best rank-$k$ approximation 
of $R$ with respect to the Frobenius norm is
\begin{equation}
||R - U^T_k \Sigma_k V_k||_F^2 = \sum_{i j} (r_{i j} - \sum_k \sigma_k u_{k i} v_{k j})^2,
\label{eq:frob}
\end{equation}
where $U_k$ is an ${m\times k}$ and $V_k$ is an ${n\times k}$ matrix containing 
the first $k$ columns of $U$ and $V$ and the diagonal $\Sigma_k$ 
containing first $k$ entries of $\Sigma$.

The \rmse\ differs from the above equation only in that summation is over known ratings
\begin{equation}
\rmse^2 = \sum_{i j\in \text{known}} \err_{i j}^2 \mbox{ where } \err_{i j} = 
r_{i j} - \sum_k \sigma_k u_{k i} v_{k j}.
\label{eq:rmse}
\end{equation}

Early works \cite{sarwar00application} used SVD for recommenders by 
defining various strategies for handling the missing values in 
the rating matrix $R$  \cite{kurucz07spectral}.  
The most natural idea is to impute the missing elements by zeroes, averages, 
or even repeatedly re-fill by predictions.
It has turned out that all above missing value imputation methods overfit 
to the imputed values \cite{kurucz07spectral}.  More recent results emphasize 
the importance of regularization to avoid 
overfitting \cite{bell2007lnp,takacs2008investigation}.
For this reason, the recommender systems community turned away from 
SVD and use other optimization methods for rating matrices 
with missing values, most notably stochastic gradient 
descent \cite{takacs2008unified} and alternating least squares \cite{koren2009matrix}.

In our problem, locality is an additional information that can be exploited 
for analyzing the recommender matrix.
Surveys on recommendations in location-based social networks 
\cite{bao2013survey,symeonidis2014location} combine spatial ratings 
for non-spatial items, nonspatial ratings for spatial items, 
and spatial ratings for spatial items \cite{levandoski2012lars}. 
Flickr geotags are used for travel route recommendation, concentrating 
on routes and not individual places in \cite{kurashima2010travel}.
User similarity based methods may combine friendship information 
with the distance of the user home locations \cite{ye2010location,ye2011exploiting}.


Most similar to our method is the Probabilistic Matrix Factorization approach
that fuses geographic information \cite{cheng2012fused} and
observes that ``users tend to
check in around several centers, where the check-in locations
follow a Gaussian distribution at each center [\ldots and]
the probability of visiting
a place is inversely proportional to the distance from
its nearest center; if a place is too far away from the location
a user lives, although he/she may like that place,
he/she would probably not go there.'' 

\section{The Nomao Datasets}

Nomao is a startup company located in France \cite{nomaocom}. 
It performs the analysis of point of interest (POI) rating and reservation services 
and collects POI information including user ratings from France
with a special accent on Paris 
and Toulouse regions where the company headquarters are located. 
The Nomao dataset used in our experiments contain user-POI ratings, 
and GPS information of the rated POIs. We investigate two separated datasets. 
The first one contains information on POIs in France, 
while the second has ratings only on POIs located in Paris.
We analyze the  datasets  collected during the time period up to year 2012. 

Table~\ref{tab:attributes} (\textbf{top}) shows the basic attributes 
of the original datasets. The average number of ratings per item is relatively large, 
the average number of ratings per user is very low. Moreover, only 
a very few percent of all user-item scores is known.

The distribution $P(\nu)$ of frequency  of votes per spot $\nu$ 
(or item $i$) for France dataset is shown in the top panel of Fig.~\ref{figtimevol}.
This distribution is stable in time and is well described by the power law
$P(\nu) \propto 1/\nu^a$ with $a \approx 1.5$. Also, the distribution
 $P(\mu)$ of frequency  of votes per customer $\mu$ (or user $u$)
remains stable in time with the power law
dependence $P(\mu) \propto 1/\mu^b$ with $b \approx 2.75$.
It is important to note that these distributions
remain stable from year 2004 up to year 2012 even
if the number of votes increases almost by two orders of magnitude
during this period.
At the moment we cannot provide theoretical
reasons for the values of these exponents.

\begin{figure}
  \centering
  \includegraphics[width=\columnwidth]{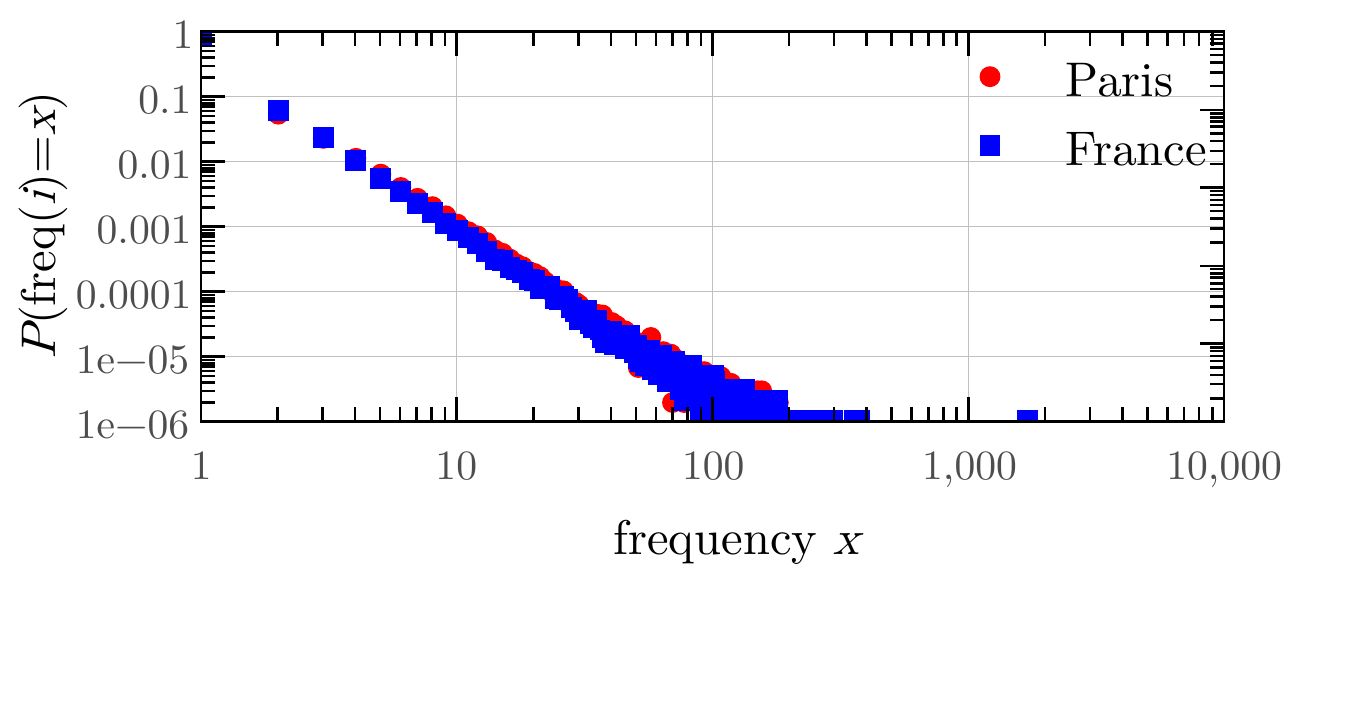}
  \includegraphics[width=\columnwidth]{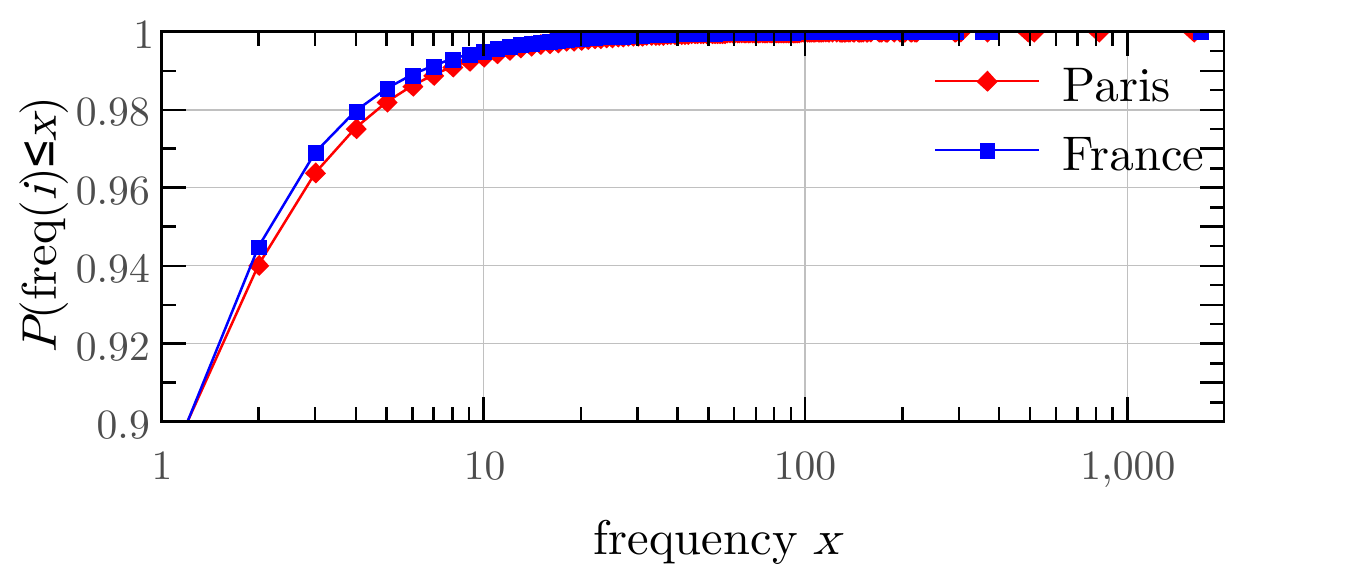}
  \caption{Probability density function (\textbf{top}), 
and cumulative density function (\textbf{bottom}) 
of user activities in the original datasets.}
  \label{fig:user_act}
\end{figure}

\begin{figure}
\centering          
    \includegraphics[width=\columnwidth]{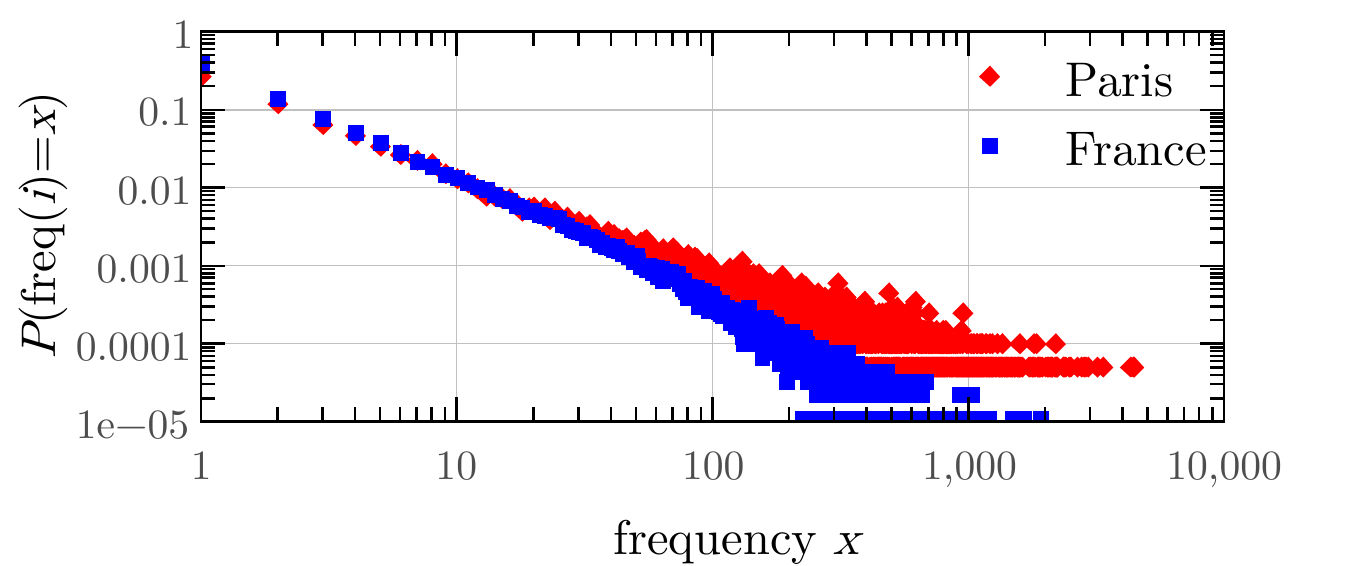}
    \includegraphics[width=\columnwidth]{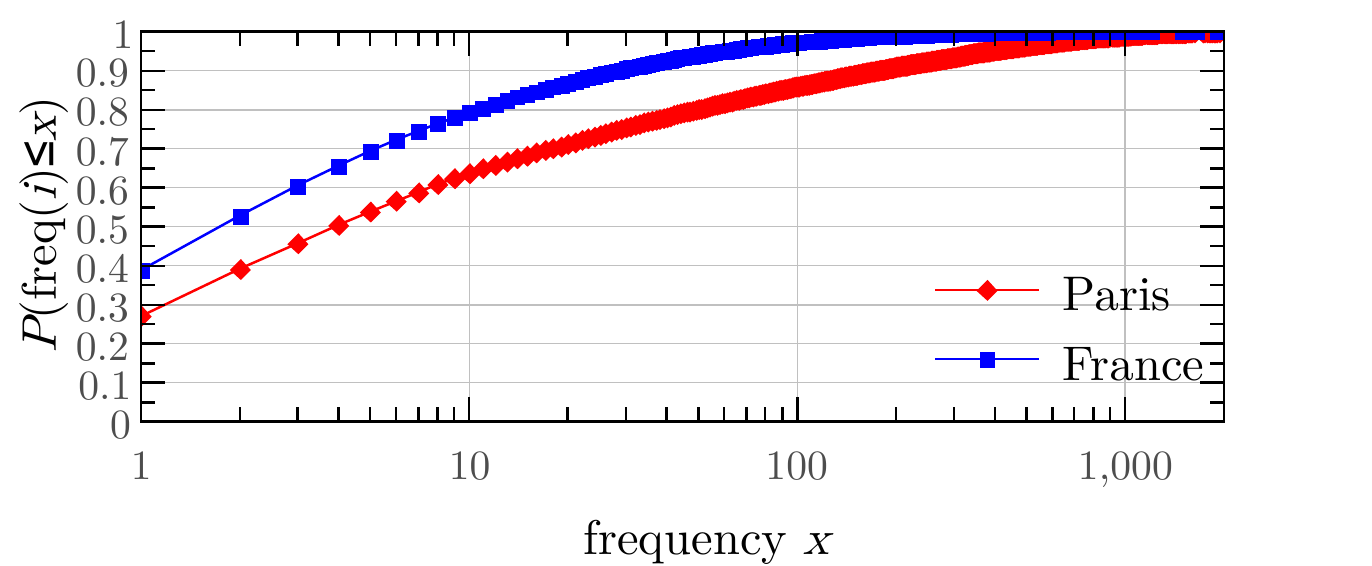}
    \caption{Probability density function (\textbf{top}), 
and cumulative density function (\textbf{bottom}) 
of item activities in the original datasets.}
    \label{fig:item_act}
\end{figure}

We call user activity how many times a user scored different items. 
We define item activity similarly. Fig.~\ref{fig:user_act} shows 
the probability density function (PDF), an the cumulative density function (CDF) 
of user activities. Fig.~\ref{fig:item_act} shows the same distributions for items. 
Both user and item activities follow power-law distributions
with the exponent values being very similar for France and Paris datasets.
As in Fig.~\ref{figtimevol} we find that the exponent
for probability of votes for POIs is $a \approx 1.5$
while the exponent for the exponent for probability of votes
of users is $b \approx 2.75$.

To handle the extreme sparsity of the user-item matrices, 
we selected a smaller subset of the user-item rating datasets 
by the following selection criteria:
\begin{itemize}
  \item We only used ratings between 0-5. Part of the ratings,
 probably originating from a different system, were out of this range.
  \item We filtered out users and items that have less than $A$ ratings. 
In other words, we selected the subgraph of the user-item rating 
bipartite graph with users and items that have degree at least $A$. 
For Paris we set $A = 10$, for France we set $A = 5$.
\end{itemize}
Table~\ref{tab:attributes} (\textbf{bottom}) shows the attributes 
of the selected subsets. In what follows we use these datasets in our experiments.

\begin{table}
    \centering
    \caption{Attributes of the original (\textbf{top}), and cleaned (\textbf{bottom}) datasets.}
    \begin{tabular}{|r||l|l|}
    \hline
    \textbf{original}                  & Paris         & France \\ \hline
    Number of ratings        & 1,539,964   & 1,432,601 \\ \hline
    Number of users              & 998,127     & 1,077,568   \\ \hline
    Number of items              & 20,576      & 99,976    \\ \hline
    Average ratings per user     & 1.543       & 1.329     \\ \hline
    Average ratings per item     & 74.84       & 14.32     \\ \hline
    Ratio of known ratings       & 0.0075\%     & 0.0013\% \\ \hline
    \textbf{cleaned}                        & Paris   & France \\ \hline
    Number of ratings        & 114,352 & 97,452\hspace{0.65cm} \\ \hline
    Number of users          & 5,756   & 9,471  \\ \hline
    Number of items          & 2,952   & 7,605  \\ \hline
    Average ratings per user & 19.87   & 10.29  \\ \hline
    Average ratings per item & 38.74   & 12.81  \\ \hline
    Ratio of known ratings  & 0.672\%  & 0.135\% \\ \hline
    Average rating          & 3.714    & 3.747 \\ \hline
    \end{tabular}
    \label{tab:attributes}
\end{table} 

\begin{figure}
    \includegraphics[width=\columnwidth]{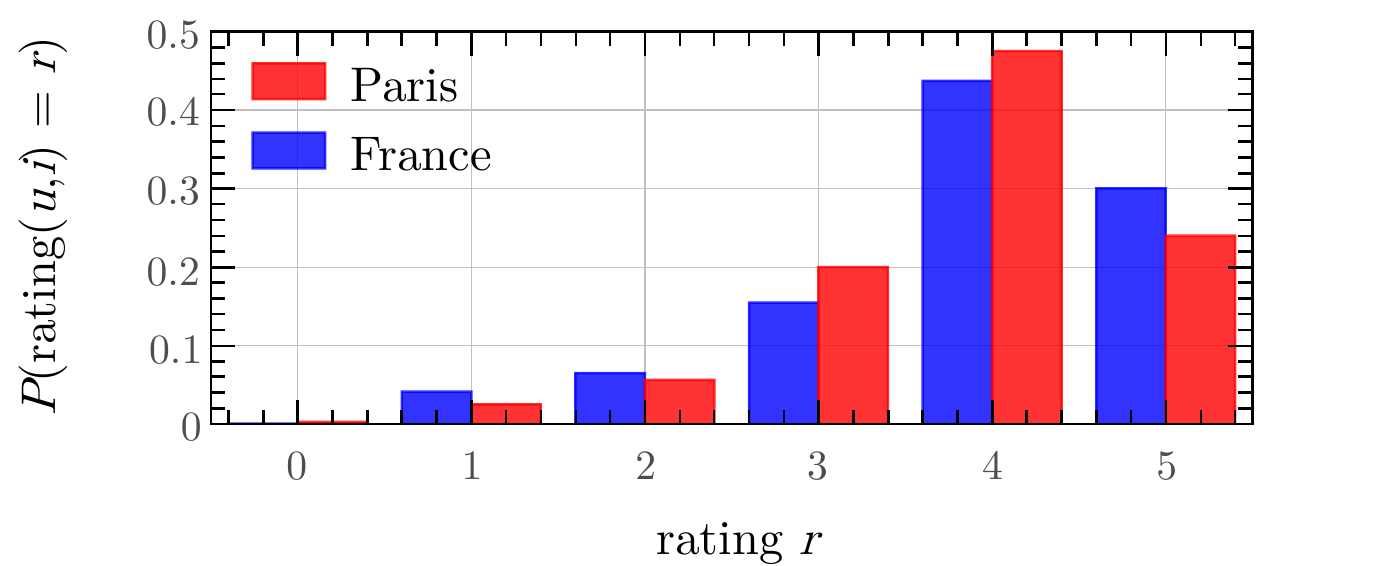}
    \includegraphics[width=\columnwidth]{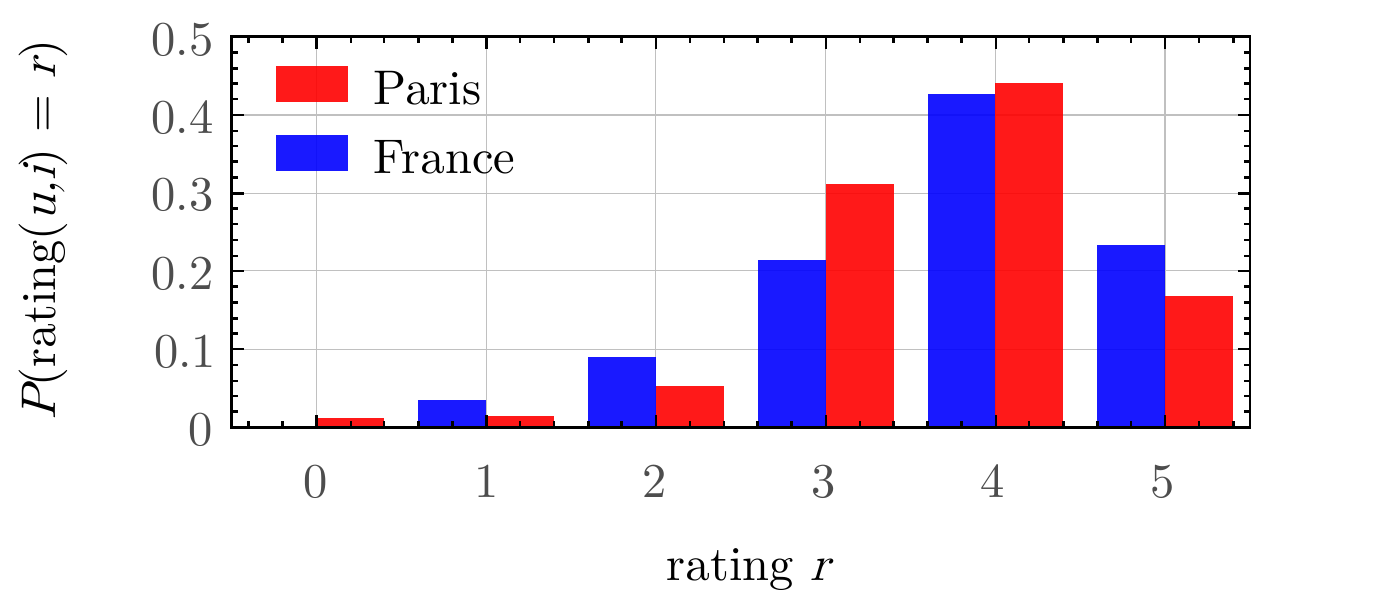}
    \caption{Score distributions with integer binning.
\textbf{ Top}: original dataset, \textbf{Bottom}: cleansed dataset.}
    \label{fig:score_dist}
\end{figure}

In Fig.~\ref{fig:score_dist} we show the score distributions: 
the top (bottom) panel shows the distributions for the original (cleaned) 
datasets. We see that the original and cleaned datasets have 
similar distributions of scores. 

\begin{figure}
          \centering
          \includegraphics[width=\columnwidth]{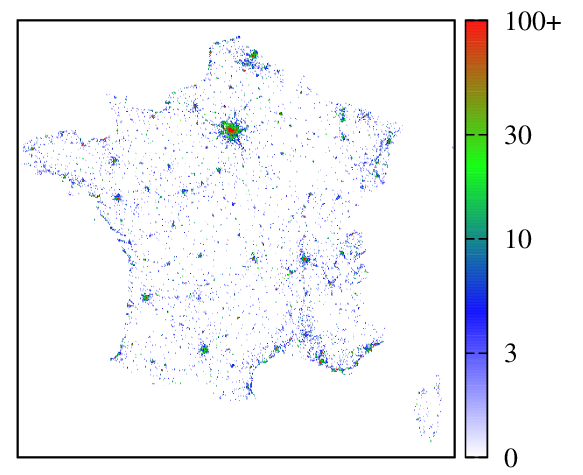}
          \includegraphics[width=\columnwidth]{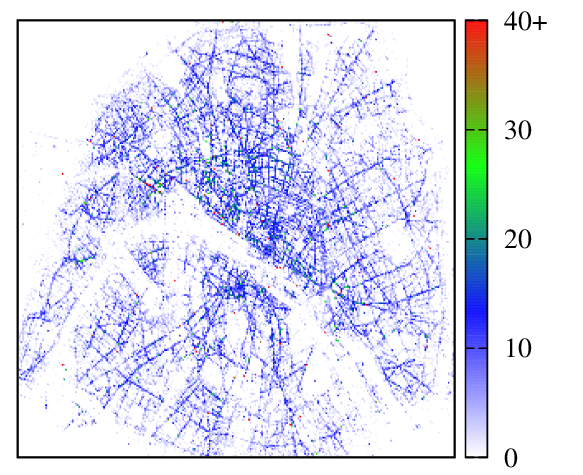}
\caption{Geographical distribution of POIs in the original datasets. 
Top panel: case of France
(each square pixel represents  $7.8 {km}^2$); bottom panel: case of Paris
(each square pixel represents  $1370  {m}^2$); 
color bars give number of POIs per pixel (cell); 
a limitation  in number of POIs 
is introduced for a better color representation.} 
\label{fig:geo_distance}
\end{figure}

Fig.~\ref{fig:geo_distance} shows the geographical density of POIs 
in France (top) and Paris (bottom) for the original datasets.
The geolocation data of POIs are used in the following Sections for
spectral analysis.

In the following we perform
all computations with the cleaned datasets since 
the analysis of multiple votes of the same user
provides more reliable statistical data.


\section{Spectra of recommender matrices}
\label{sect:reco-matrix-spectrum}

\subsection{Singular value decomposition}

The recommender matrix $R$ consists of the preference values $r (u,i)$ 
of users $u$ for items $i$.  The values may denote \emph{explicit} rating values, 
e.g.\ 1-5 stars for Netflix movies \cite{bell2007lnp}. We may also consider 
the so-called \emph{implicit} ratings problem, where the value is 1 if 
the user visited POI $i$ and 0 otherwise.  
The value of the explicit matrix is missing whenever the user gave no rating yet.
In most of the cases, this matrix is very sparse with only 1\% or less known values.
The implicit matrix is always a full 0--1 matrix, 
however the 0 values are uncertain: the user may not know about 
the item or had no time yet to visit it.

The so-called \emph{Latent Factor Model} is an approximation 
$\hat{R}$ of the original rating matrix $R$,
\begin{equation}
\hat{r} (u, a) = \sum_{f=1}^k p_{u f} q_{a f},
\end{equation}
where $P = [p_{u f}]$ and $Q = [p_{u f}]$ are 
the user and item factor models, respectively.

For a fixed number of factors $k$, $\hat{r}$ approximates $r$ 
with the smallest root mean squared error if it is defined by 
the singular vectors corresponding to the $k$ largest singular values,
\begin{equation}
\hat{r} (u, a) = \sum_{f=1}^k p_{u f} q_{a f},
\end{equation}
where the singular value decomposition (SVD) of $R$ is $U \Sigma V^T$.

Since
\begin{equation}
R R^T = U \Sigma^2 U^T \mbox{ and } R^T R = V \Sigma^2 V^T,
\end{equation}
the spectrum of the recommender matrix $R$ is defined identically by 
the square root of the eigenvalues of $R R^T$ or $R^T R$.  
These latter matrices are symmetric positive semidefinite, 
the spectrum is non-negative real.

If $R$ contains missing values such as in the case of an explicit rating matrix, 
SVD is undefined. We may still define the best root mean square approximation 
by summing the error for the known ratings only as in equation \eqref{eq:frob}.

\subsection{Stochastic gradient descent: Latent factor modeling with missing values}

We use the regularized matrix factorization method of \cite{takacs2008investigation}
and optimize the minimum squared error of the $k$-factor model
\begin{equation}
\hat{r} (u,i) = \sum_{l=1}^k p_{u l} q_{i l},
\end{equation}
where $p$ and $q$ contain the user and item models, respectively.
By adding regularization with weight $\lambda$, we optimize the quantity
\begin{equation}
\resizebox{0.9\hsize}{!}{$\sum_{u,i} \left(r (u,i) - \sum_{l=1}^k p_{u l} q_{i l}\right)^2 + 
\lambda\sum_u\sum_{l=1}^k p_{u l}^2 + \lambda\sum_i\sum_{l=1}^k q_{i l}^2.$}
\end{equation}
For a single event $(u,i)$ we optimize the coefficients $p_{u l}$ and $q_{i l}$ 
for $l=1$, $\ldots$, $k$ by gradient descent with learning rate lrate as
\begin{eqnarray}
\resizebox{0.9\hsize}{!}{$p_{u l} \leftarrow p_{u l} + \mbox{lrate} \cdot
 \left(r (u,i) - \sum_{l=1}^k p_{u l} q_{i l}\right)q_{i l} - \mbox{lrate} \cdot \lambda p_{u l};$}\\
\label{eq:sgd-p}
\resizebox{0.9\hsize}{!}{$q_{i l} \leftarrow q_{i l} + \mbox{lrate} 
\cdot \left(r(u,i) - \sum_{l=1}^k p_{u l} q_{i l}\right)p_{u l} - \mbox{lrate} \cdot \lambda q_{i l}.$}
\label{eq:sgd-q}
\end{eqnarray}

Unlike SVD where eigenvalues are sorted, the SGD factors are not ordered by the above equations. 
 In order to produce the eigenvector maps, we built ranked factors 
by an iterative SGD that optimize only on a single factor at a time \cite{funk06netflix}.

\subsection{Mapping SVD and SGD latent factors}

\begin{figure} [!ht]
\centerline{\includegraphics[width=6cm]{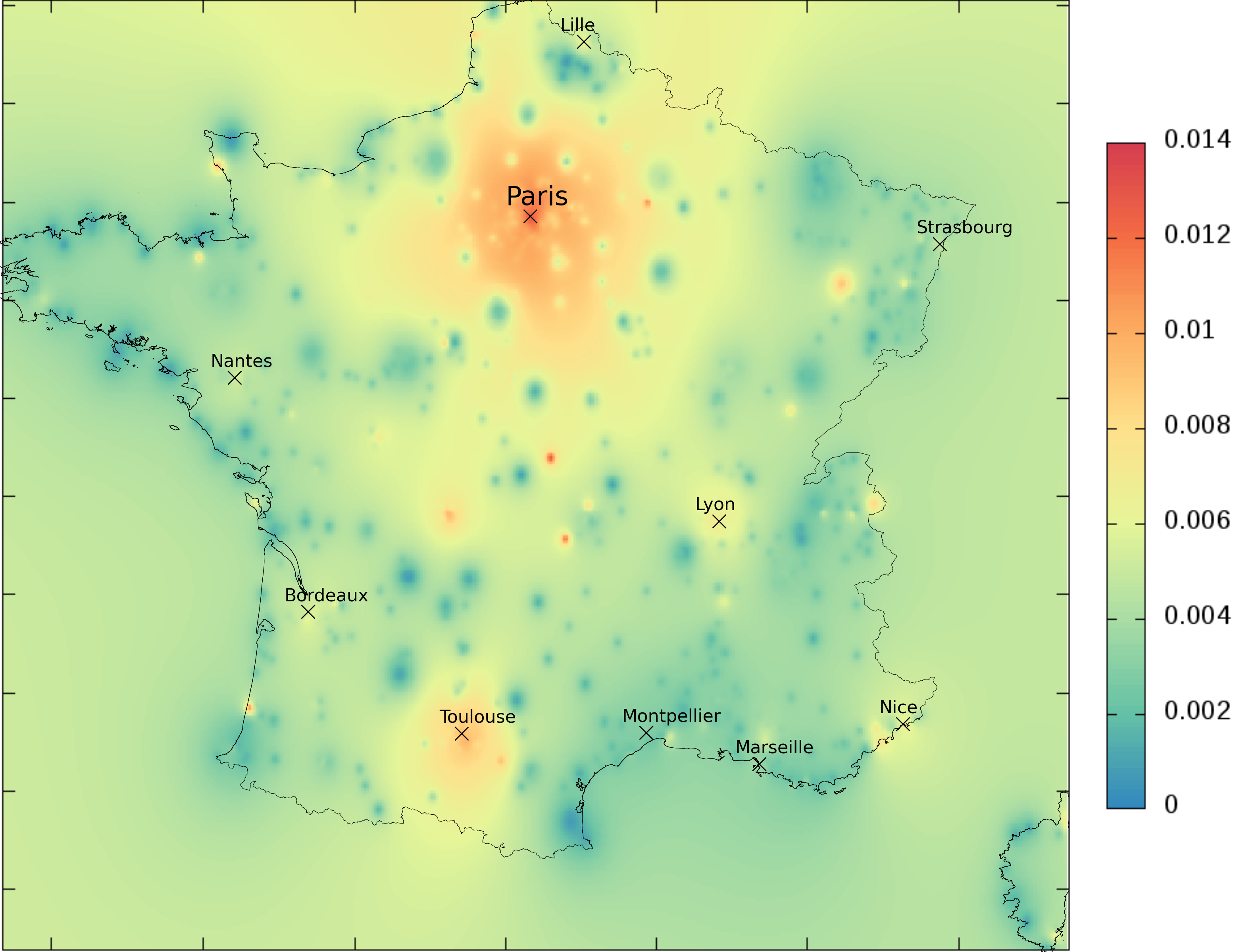}%
\includegraphics[width=6cm]{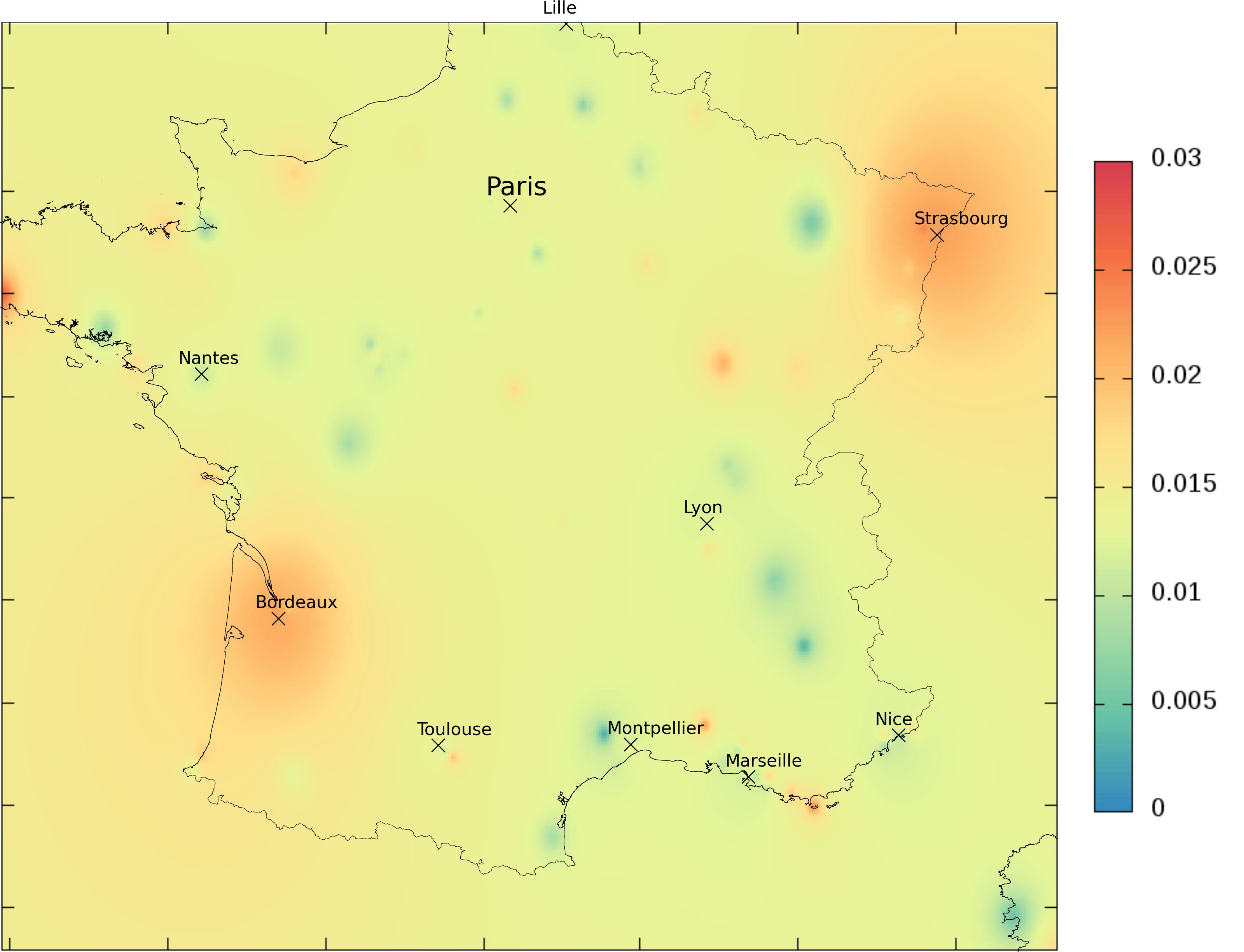}}
\vspace{1cm}
\centerline{\includegraphics[width=6cm]{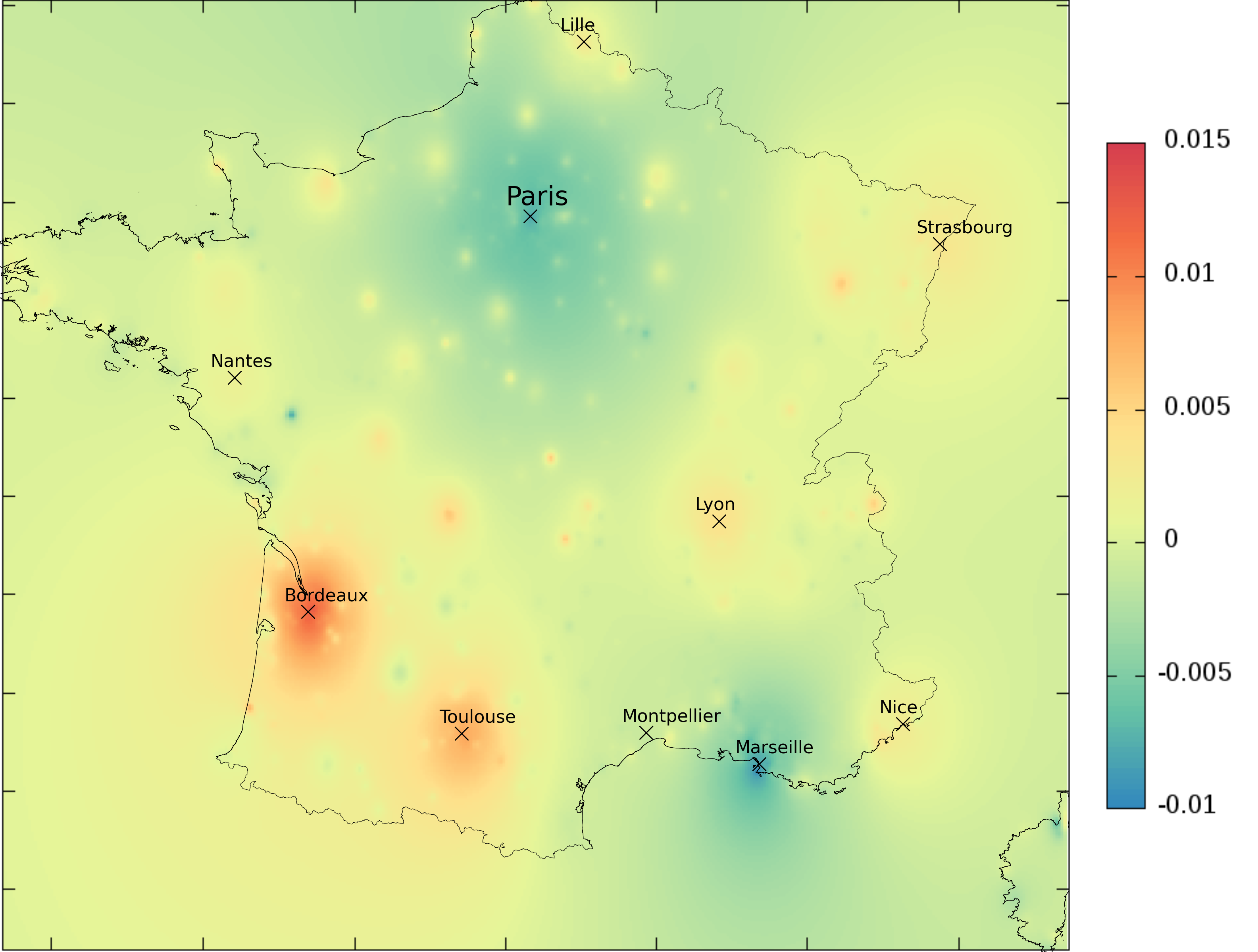}%
\includegraphics[width=6cm]{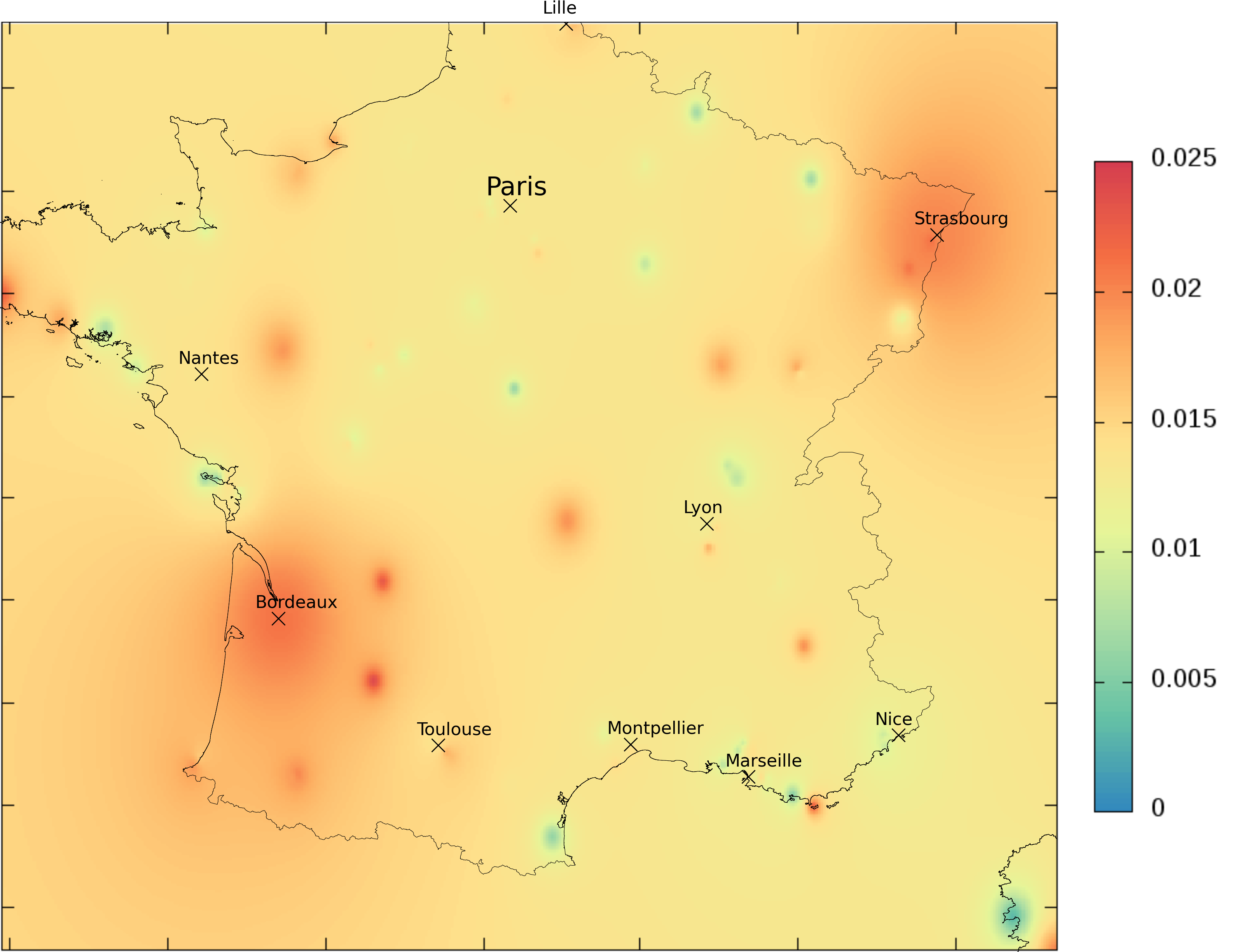}}
\vspace{1cm}
\centerline{\includegraphics[width=6cm]{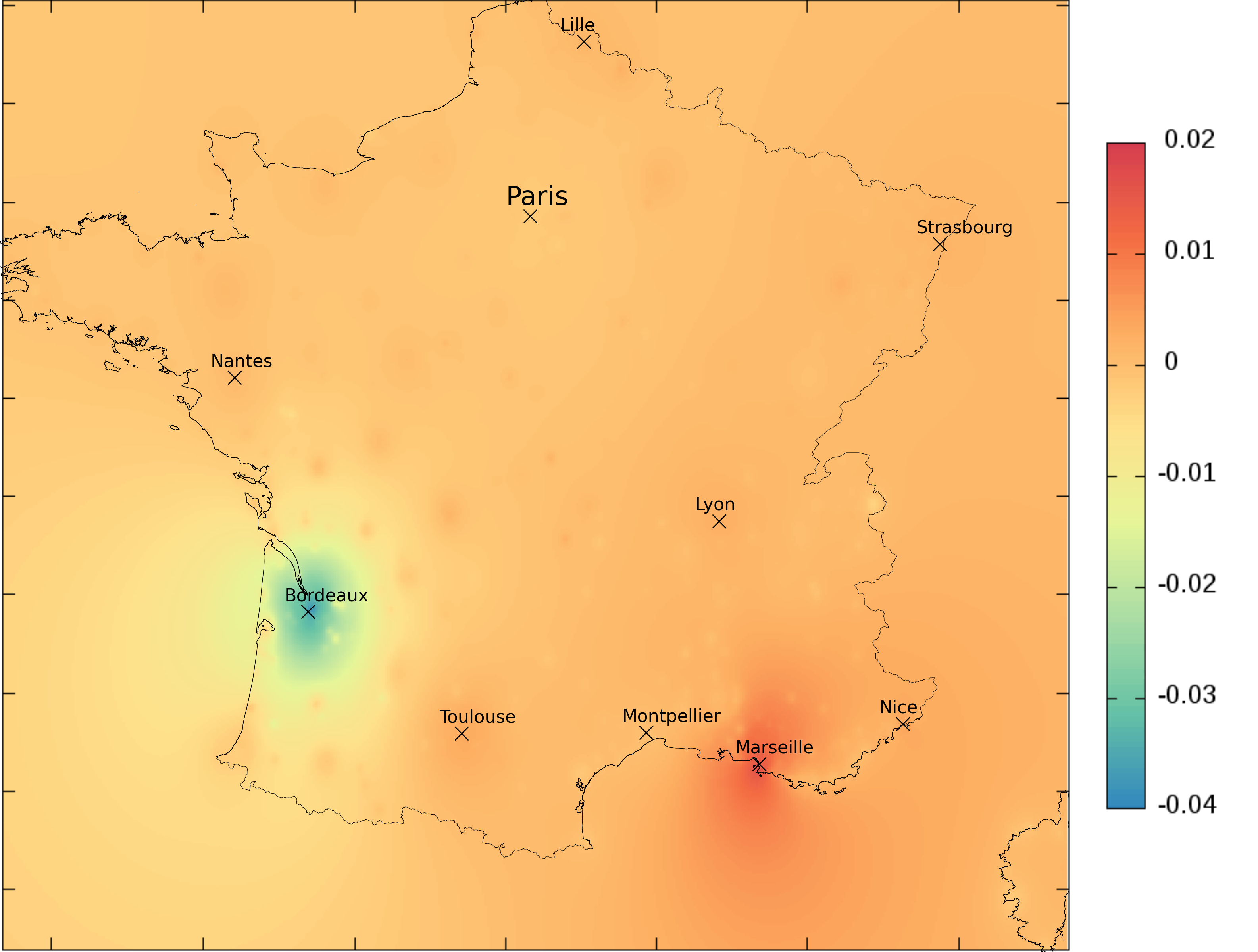}%
\includegraphics[width=6cm]{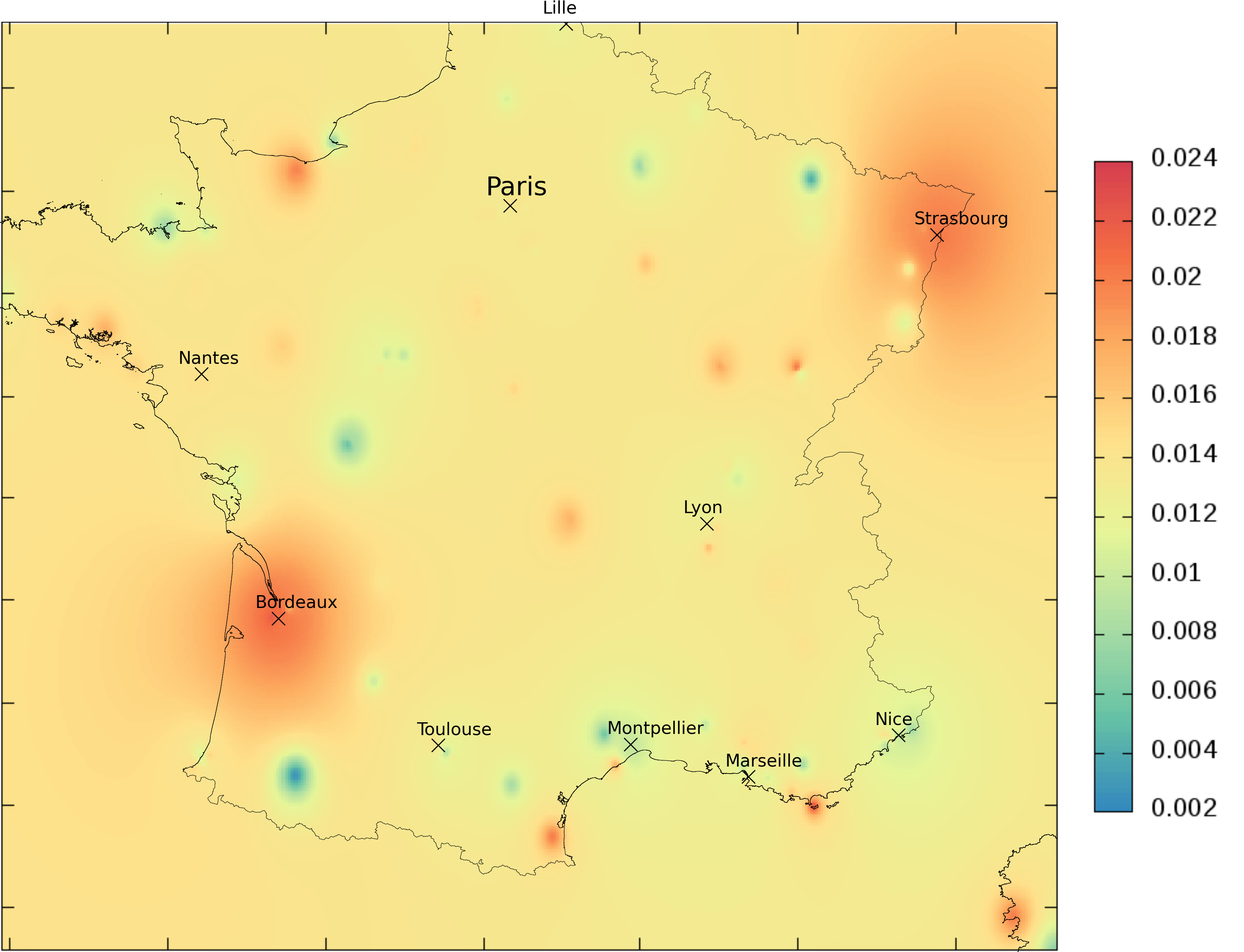}}
\caption{The first, second and fourth singular vectors of the Nomao France rating matrix 
by SVD (\textbf{left}) and SGD (\textbf{right}). Here, the SVD eigenvectors correspond 
to Paris and Toulouse; Bordeaux, Toulouse and Marseille; Bordeaux and Marseille
 respectively, 
while the SGD plots for respective vectors are scattered around several cities.}
\label{fig:eigenvectors} 
\end{figure}

First we set each unknown value of $R$ to zero and computed the SVD decomposition. 
The first, second,  and fourth singular vectors are plotted over 
the map of France (Fig.~\ref{fig:eigenvectors}, left) by assigning the value 
in the vector to the location of the POI. More specifically, we averaged these 
values on a grid to create the final heatmaps. The smoothing algorithm weighted 
the value of each POI to the closest grid point inversely proportional 
to their euclidean distance.

The heatmaps in Fig.~\ref{fig:eigenvectors}, left, indicate that the singular 
vectors are strongly geolocation related. The first few dimensions correspond 
to the largest cities in France.

Similarly, we investigated the latent vectors of $R$ computed with the SGD algorithm. 
The first, second and fourth latent vectors are plotted over the map of France 
in Fig.~\ref{fig:eigenvectors}, right, similar to the SVD eigenvectors. 
While the SVD singular vectors were centralized one-by-one on a large city, 
the SGD latent factors are the linear combination of them. The latent factors 
are also geolocation related, but not separated among 
the main cities like the SVD singular vectors.

\begin{figure} 
\centering
\includegraphics[width=\columnwidth]{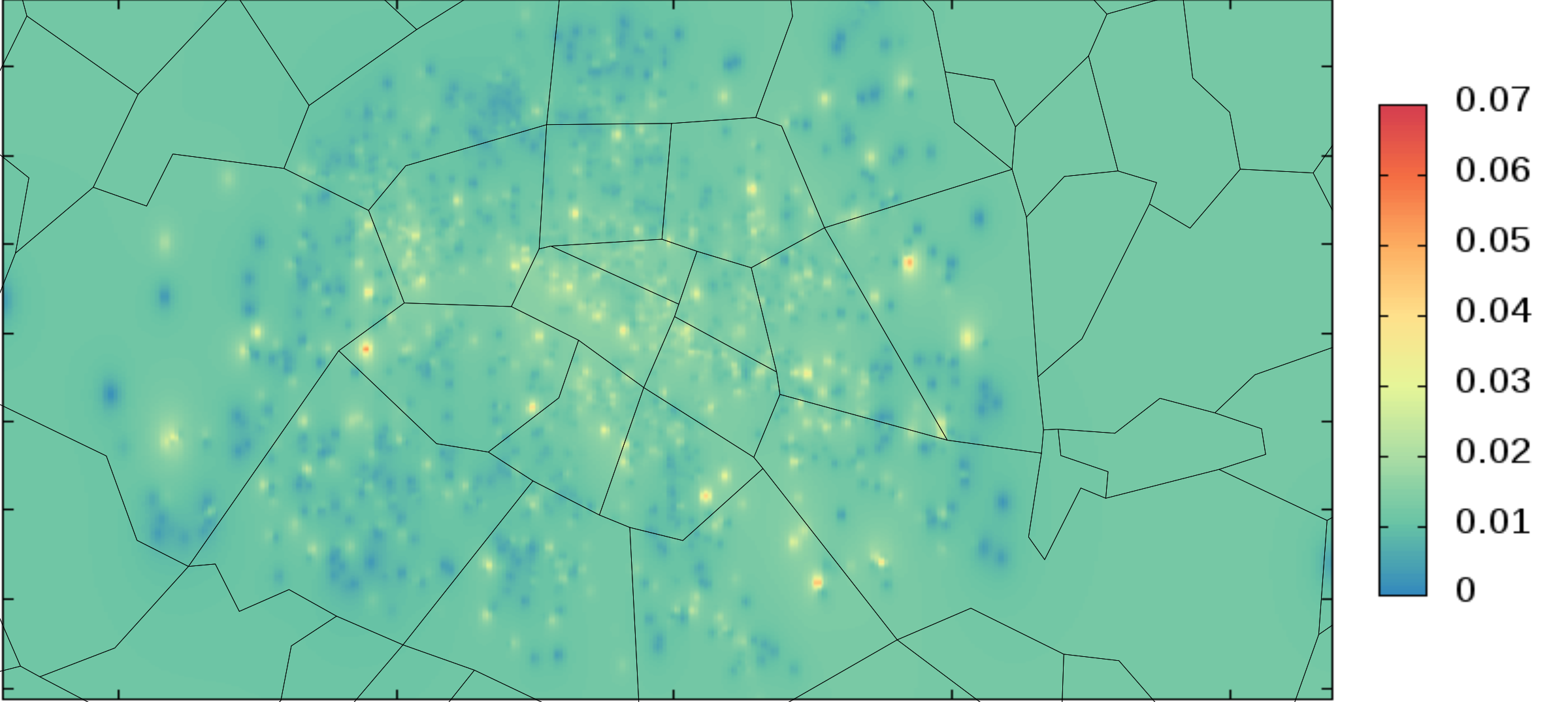}\vspace{0.5cm}
\includegraphics[width=\columnwidth]{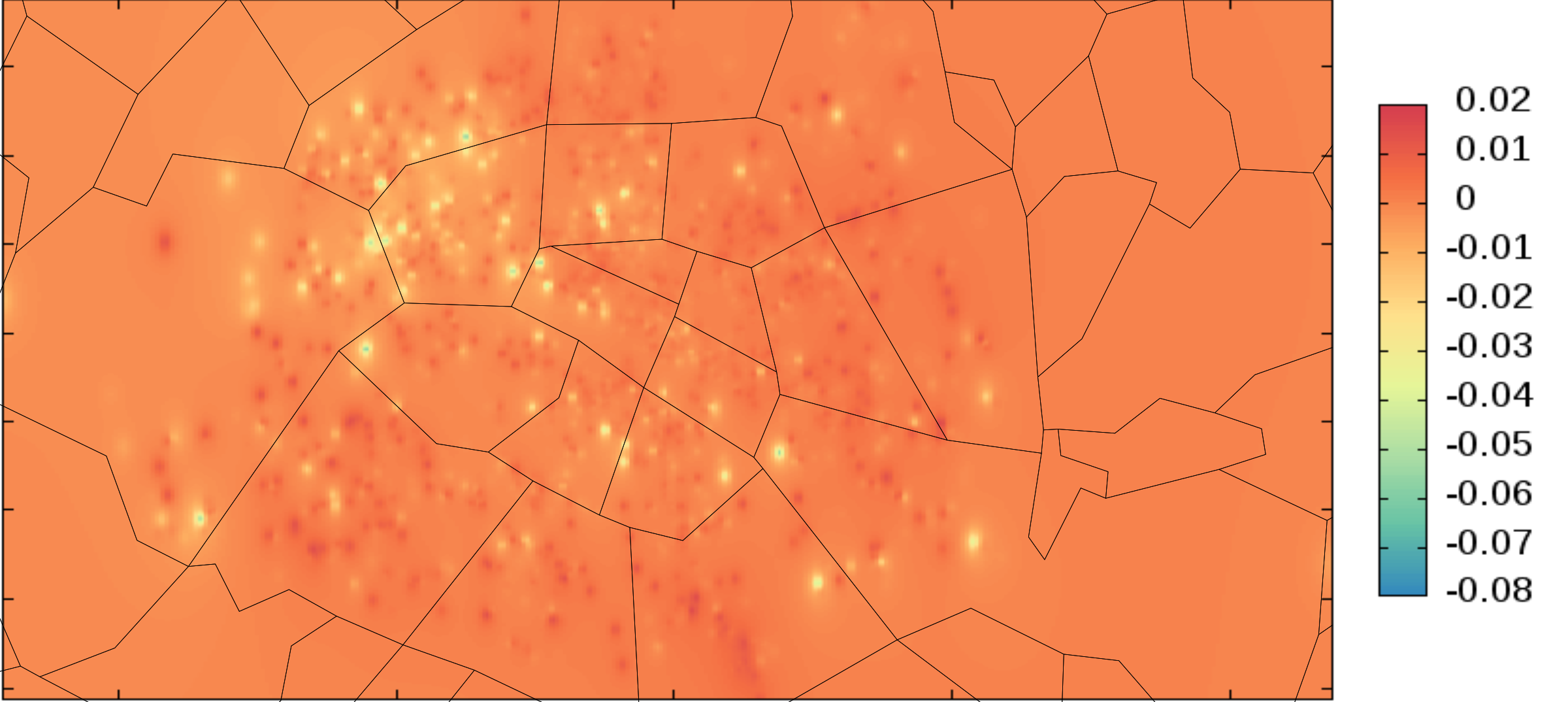}\vspace{0.5cm}
\includegraphics[width=\columnwidth]{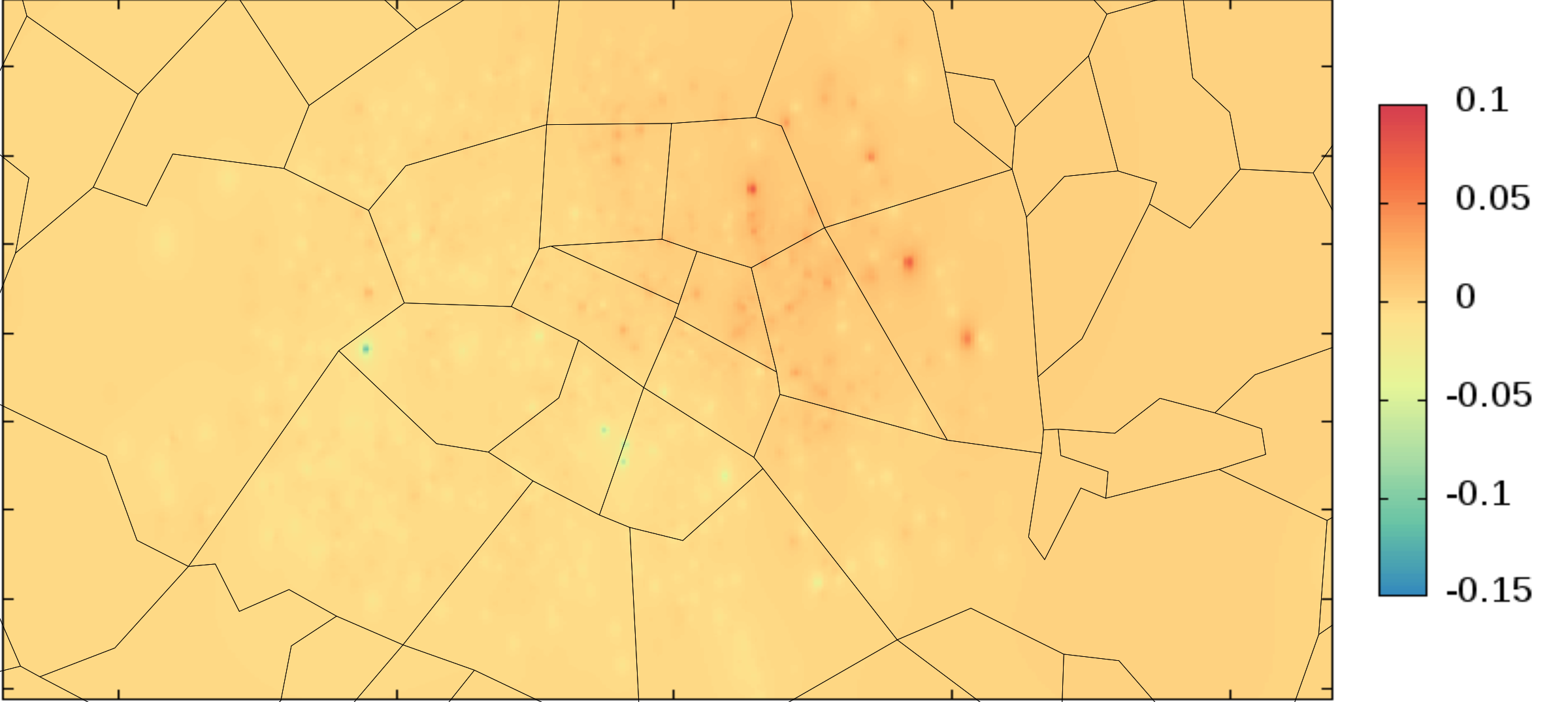}
\caption{The first, third and fourth singular vectors of 
the Nomao Paris rating matrix obtained by SVD.}
\label{fig:svd-paris}
\end{figure}

In \ref{fig:svd-paris} we mapped the first three singular vectors of the Paris dataset. 
The different vectors may focus on different districts. However, they are not 
as clearly separated as the singular vectors of the France dataset.

In Section~\ref{sec:collaborative_filtering} we use these key observations 
to improve the recommendation quality of the SGD.

\newpage
\onecolumn

\newpage
\twocolumn

\section{Prediction for ratings and visits to locations}
\label{sec:collaborative_filtering}

\subsection{Recommender evaluation}
\label{sec:recommender_eval}

Recommender systems serve to find \emph{new} products for the users 
that are relevant for them. More specifically, for a given user $u$ 
an item~$i$ a recommender system may retrieve the predicted relevancy $\hat{r}_{ui}$. 
This is called the rating prediction task. The Netflix Prize competition 
\cite{netflix-prize} was a challenge in rating prediction.  
While in the Netflix Prize competition, contestants were optimizing 
to predict all ratings to the users, a recommender system in practice selects 
the top rated items for a given user.  In this top-$K$ prediction task 
\cite{deshpande2004item,cremonesi2010performance,yuan2011factorization}, 
a recommender system should retrieve for a given user a top list of items with length $K$. 
The top list should contain the most relevant items for the given user. 
This problem is more application related than the rating prediction task. 
In our experiments we examine both problems on the NOMAO datasets.

In addition to RMSE defined by equations \eqref{eq:frob} for full and \eqref{eq:rmse} 
for partial matrices, we use two measures that evaluate 
the accuracy of the top-$K$ recommendation task.

Recall at $k$ is defined as the number of relevant POIs among 
the highest $k$ values of row $u$ in the matrix approximation,
\begin{equation}
\mbox{Recall}_u(k) = \frac{1}{|R_u|} \sum_{i=1}^k \mbox{rel}_{u,i},
\end{equation}
where $\mbox{rel}_{u,i}$ is the actual relevance of POI $i$ for user $u$ 
in the evaluation data, and $R_u$ is the number of relevant items 
for user $u$ in the dataset. We may average for all users to obtain
\begin{equation}
\mbox{Recall}(k) = \frac{1}{|U|} \sum_u \mbox{Recall}_u(k).
\end{equation}
Normalized Discounted Cumulative Gain at $k$ weights the relevance 
by the order of the predicted values as
\begin{equation}
\mbox{NDCG}_u(k) = \frac{\mbox{DCG}_u(k)}{\mbox{iDCG}_u(k)},
\end{equation}
where
\begin{equation}
\mbox{DCG}_u(k) = \sum_{i=2}^k \frac{\mbox{rel}_{u,i}}{\log_2(i+1)}
\end{equation}
and
\begin{equation}
\mbox{NDCG}(k) = \frac{1}{|U|} \sum_u \mbox{NDCG}_u(k).
\end{equation}

In our experiments, we randomly cut the data to training and test sets. 
We only use records in the training set to set the parameters of our model. 
The lower MSE, and the higher NDCG and recall we measure on the test set, 
the better is our model.

\subsection{The rating prediction task}

\begin{figure}
\centering
\includegraphics[width=\columnwidth]{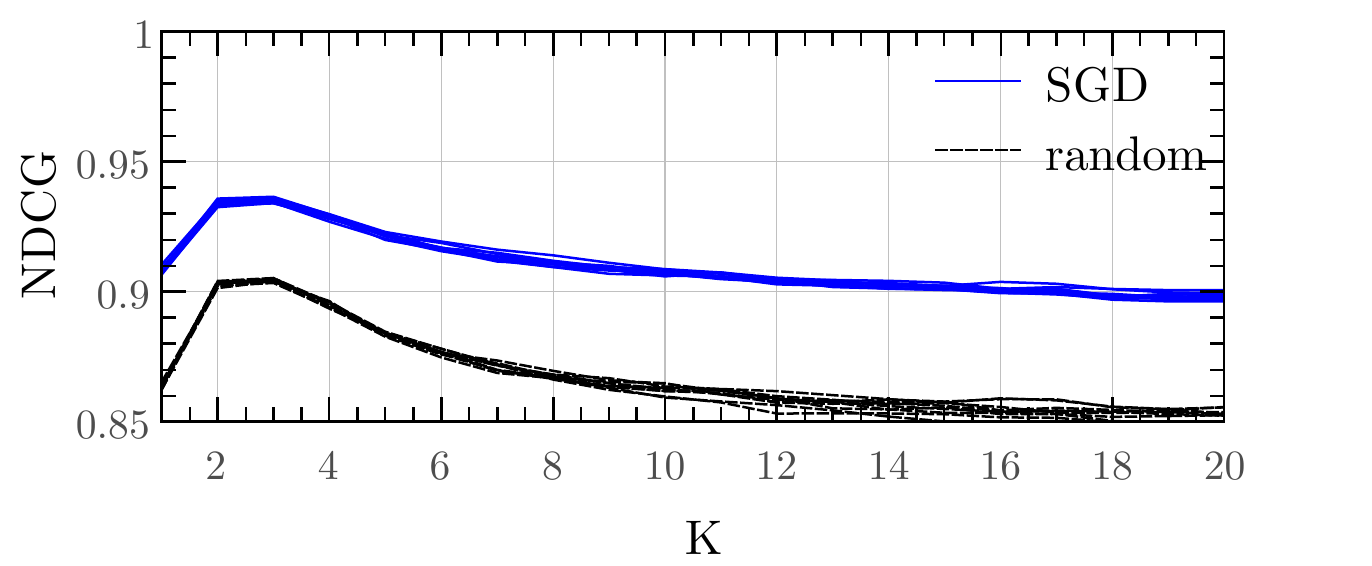}
\includegraphics[width=\columnwidth]{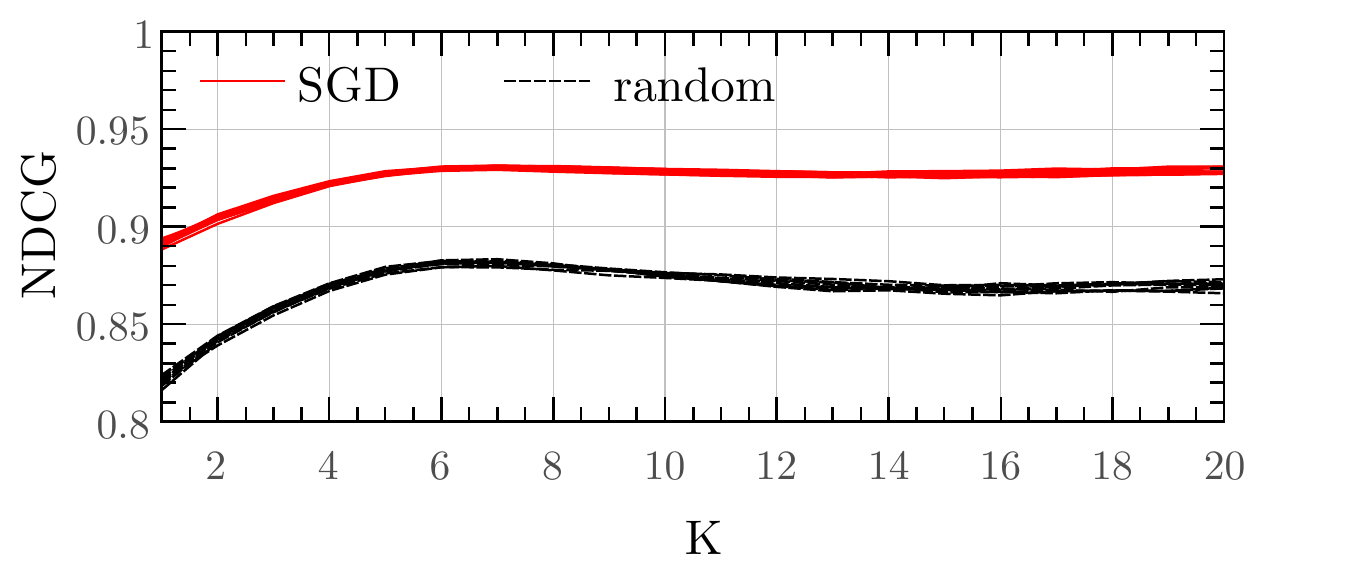}
\caption{Performance of score prediction on the  France (\textbf{top}) 
and Paris (\textbf{bottom})  datasets.}
\label{fig:pred_score_france}
\end{figure}

As indicated in Table~\ref{tab:attributes}, bottom, and in Fig.~\ref{fig:score_dist}, 
the scores have a peaked distribution. This indicates first that the rating prediction 
task makes less sense with these datasets. We trained up an SGD recommender 
by using 50\% of the datasets and computed NDCG(k) for $K=1\dots20$. 
To understand the performance of the model, we also measured the performance 
of a random recommender that predicts ratings uniform randomly. 
We repeated our experiments 10 times with 10 different random training and test sets. 
Fig.~\ref{fig:pred_score_france} shows the computed ten performance curves for 
the SGD and the baseline random recommendation. Both for SGD and the random prediction 
the ten curves are similar. This indicates the stability of our algorithms and 
evaluation metrics. We achieved significantly better result with the SGD recommender. 
However for the random algorithm, the baseline NDCG is around 0.85. 
This is due to the fact that most of the ratings are around 
the mean as the score distribution is peaked.

\subsection{Improving top recommendation with rating imputation}

Instead of simply recommending locations near to already visited places, 
we expand the training set by relying on the locality of the ratings.  
We compare our results by using SVD or SGD both for the rating matrix 
and for simply predicting the visits, i.e.\ the existence of a rating 
regardless of its value. When considering locality, 
we may identify the nearest neighbors 
by taking the absolute distance and possibly correcting by density: 
in an area densely served by POIs, customers may reach more locations, 
on the other hand, the speed of travel is likely lower than in rural areas.

For our imputation methods, let $E$ be the set of known ratings and $N_j$ 
the neighbors of location $j$. We modify the training set as follows. For all $(u,i)$,

\begin{equation}
\resizebox{0.88\hsize}{!}{
$\hat{r}_{u,i}=
\begin{cases}
r_{u,i} &\text{if $(u,i)\in E$}\\
f (R_u, N_{u,i}) &\text{if $(u,i) \notin E$ and 
for some $j$, $(u,j)\in E$ and $i\in N_j$}\\
\text{missing}&\text{otherwise,}
\end{cases}$
}
\end{equation}
where $f$ is function of $R_u$, the set of known ratings 
by user $u$, and $N_{u,i}$, the set locations visited by $u$ in the neighborhood of $i$.

In our model, we expand the list of locations per user with the neighbors of visited
places by the two strategies:
\begin{description}
\item[Constant:]
\begin{equation}
f(R_u,N_{u,i}) = c
\label{eq:const}
\end{equation}
\item[Ratings Average:]
\begin{equation}
f(R_u,N_{u,i}) = \frac{1}{|N_{u,i}|} \sum_{j \in N_{u,i}} r_{u,j}
\label{eq:rating-average}
\end{equation}
\end{description}

\begin{figure}
\centerline{\includegraphics[width=\columnwidth]{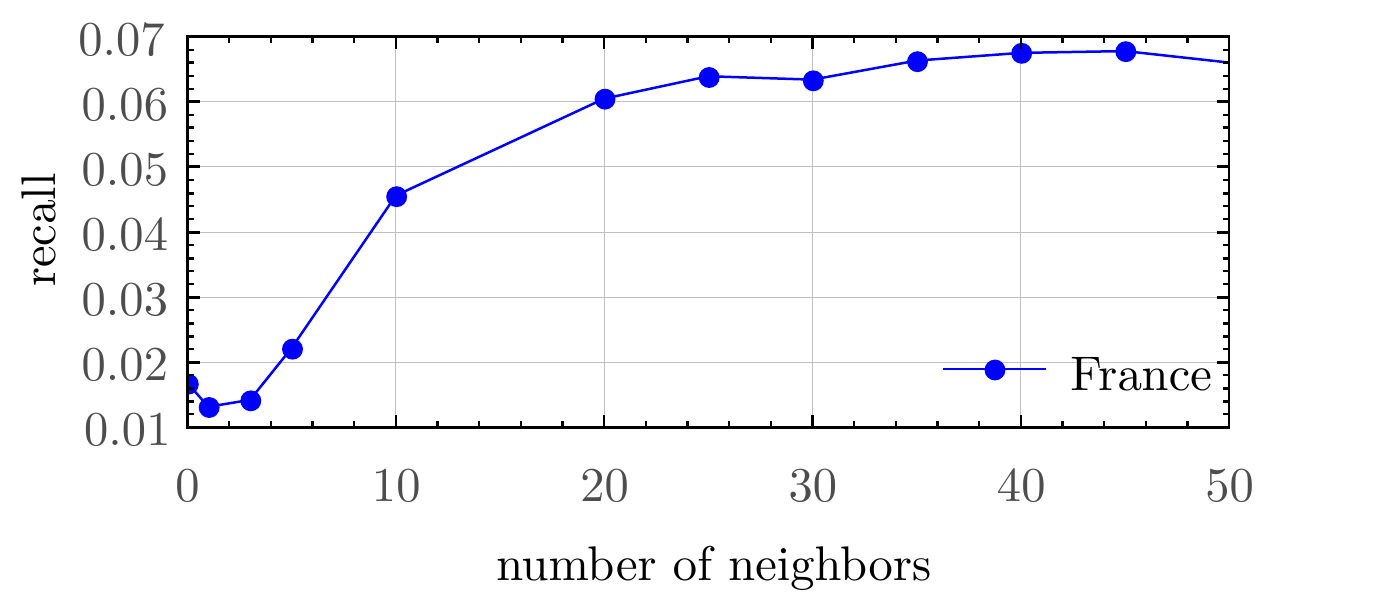}}
\centerline{\includegraphics[width=\columnwidth]{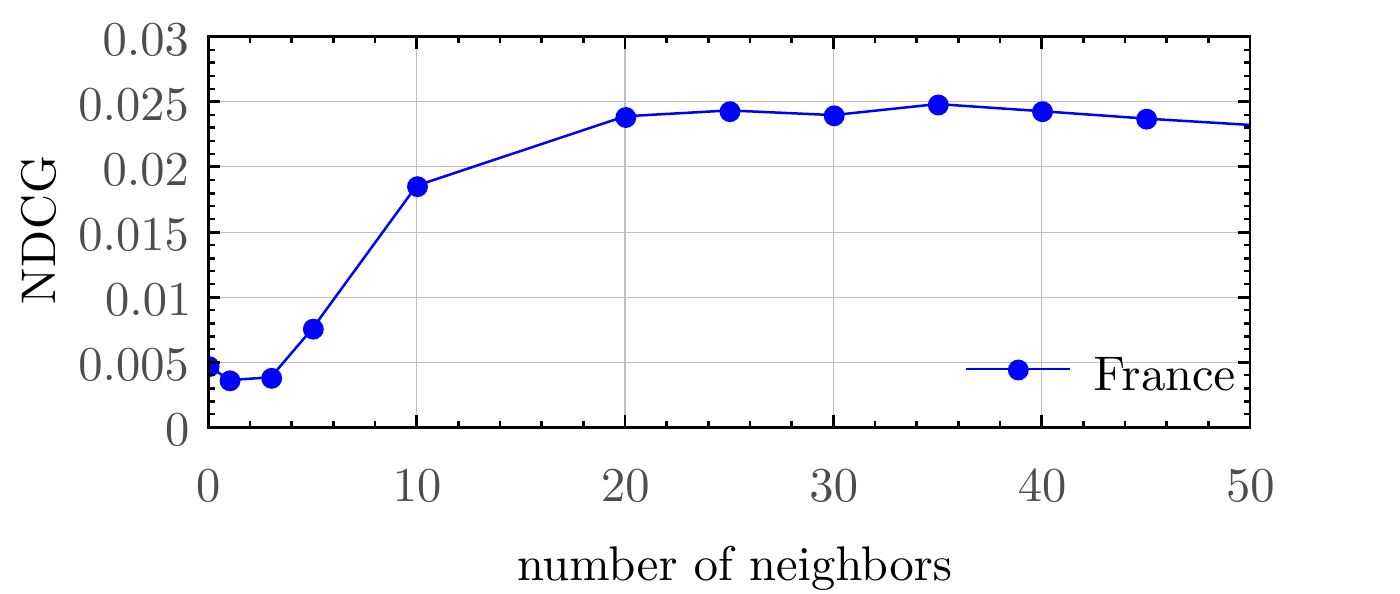}}
\caption{Recall@100 and NDCG@100 for expansion with the original ratings.}
\label{fig:ratings-expansion}
\end{figure}

The performance for expansion with the original ratings (see~\eqref{eq:rating-average}) 
on the France dataset is seen in Fig.~\ref{fig:ratings-expansion} where we observe 
that expansion by the 30-40 nearest POIs improves significantly the matrix approximation by 
the first few eigenvectors.

\begin{figure}
\centerline{\includegraphics[width=\columnwidth]{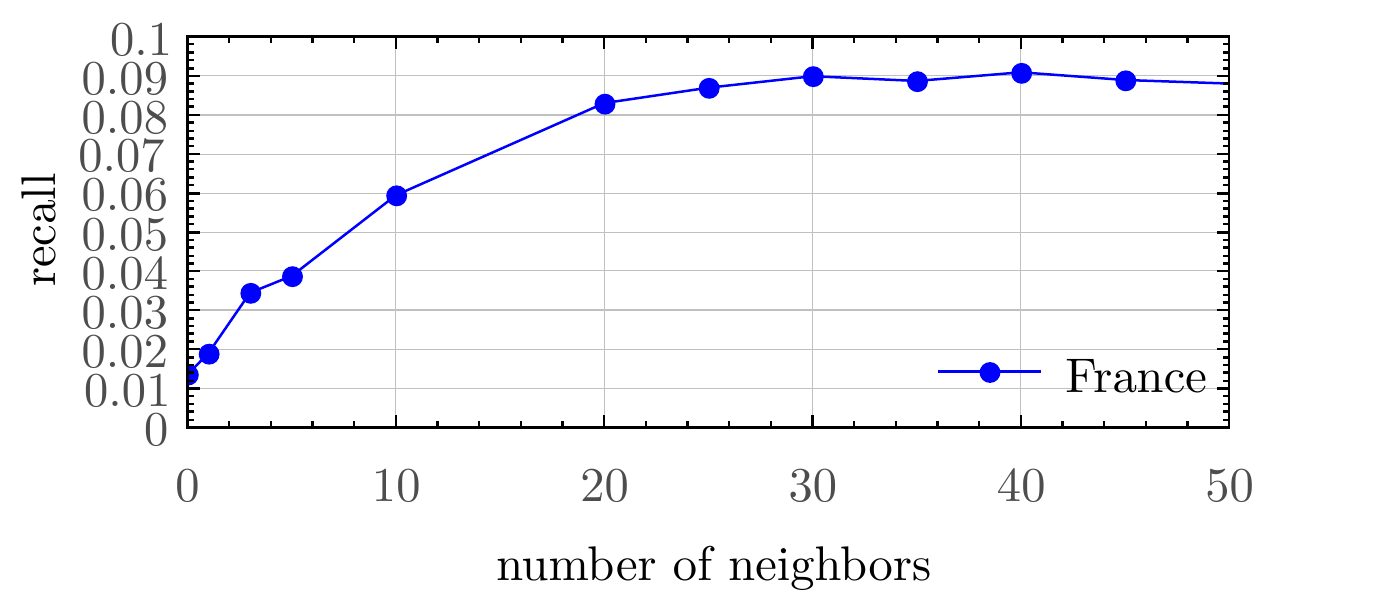}}
\centerline{\includegraphics[width=\columnwidth]{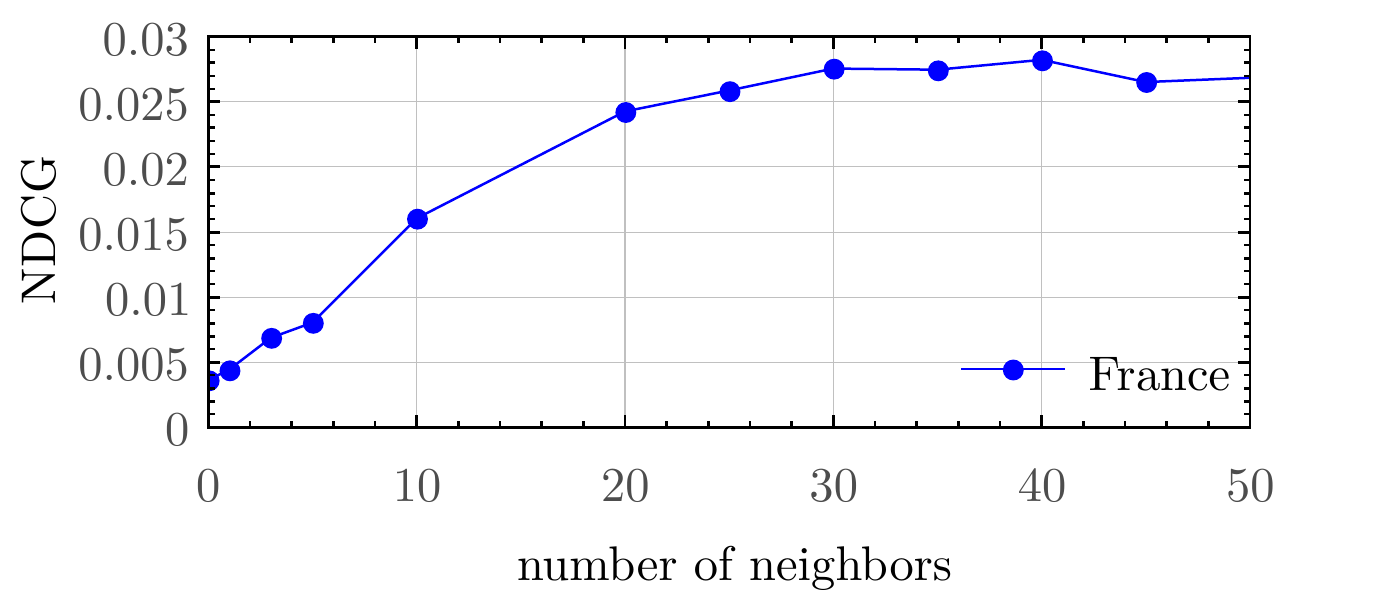}}
\caption{Recall@100 and NDCG@100 for expansion nearby visited locations.}
\label{fig:implicit-expansion}
\end{figure}

We may also consider the task of predicting which POIs the user will visit, 
regardless of the actual rating given by the user. In this so-called implicit 
recommendation task, we consider a 0--1 matrix.  Although the matrix is fully known, 
the meaning of a ``1'' is certain while a ``0'' may simply mean that the user
 has not yet had a chance to visit the POI or does not even know about it.  
Based on~\eqref{eq:const}, the performance of the implicit task with expansion 
for the France dataset is seen in Fig.~\ref{fig:implicit-expansion}
showing an improvement compared to Fig.~\ref{fig:ratings-expansion}.
However, for the Paris dataset, 
both in case of the ratings and implicit expansion experiments, we could not improve 
further the original SVD. This can be due to the fact that 
the Paris dataset is more dense geographically.


\subsection{Improving recommendation with fixed factors}

Results of Fig.~\ref{fig:eigenvectors} indicate that while SGD finds 
the most important cities in France, it can not separate them precisely. 
Furthermore, not recommending to a user POIs, that he/she have not visited,
can be easily implemented without using SGD. Indeed, SGD should learn the taste 
of the different users like in case of the movie prediction task of Netflix. 
To fix this issue in the France dataset, we selected the top $t$ cities in France. 
For a given item, we fixed the first $i$th factor to 1, if the item is located 
in the $i$th city, and 0 otherwise. We set the user factors similarly according 
to the places visited by the user in the test set. We then trained a $k$ dimensional 
latent factor model where we updated only the remaining $k-t$ dimensions. 
We compared this recommender with a traditional $k$ dimensional SGD recommender. 

\begin{figure}
\centering
\includegraphics[width=\columnwidth]{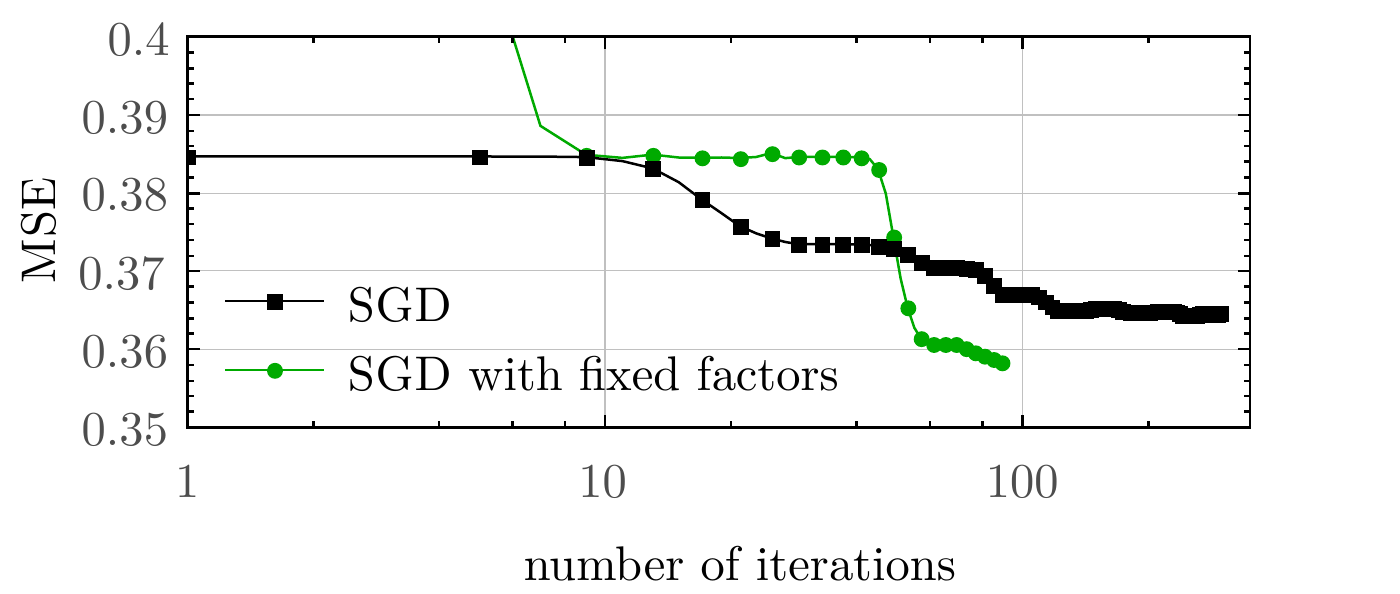}
\caption{Improved MSE results with fixed factors on the France dataset.}
\label{fig:fixed_factors}
\end{figure}

In our experiments we included the top $t=10$ cities, in order Paris, 
Marseille, Lyon, Toulouse, Nice, Nantes, Strasbourg, Montpellier, Bordeaux, and Lille.
Fig.~\ref{fig:fixed_factors} shows the MSE on the test set 
as the function of the number of iterations on the training set. 
With the fixed factor model we can achieve significantly better MSE. 
Furthermore, our best result is achieved with half less iterations 
compared to the number
needed to train the original latent factor model.

\section{Discussion}

Our statistical analysis of NOMAO votes of customers for spots of France shows
that it is described by a power law frequency distributions
with exponents $a \approx 1.5$ (for spots) and $b \approx 2.75$
(for customers) which remain stable in time
even is the number of votes is increased almost by two orders of magnitude during this
time period. Further studies are required to
establish the physical origins of such laws and to clarify
for universal they are.  
It is possible that the physical reasons for emergence of such type
distributions have certain similarities with the
phenomenon of self-organized criticality
broadly discussed in physical systems
(see e.g. \cite{bak,wikiabelsand,wikiselfcrit}).
It is interesting to note that the exponent
of cluster distribution in self-organized critical models
in 3D has an exponent close to $1.4$ \cite{bak}
being not so far from the exponent $a=1.5$ we find for
spots.

We  explored the spectrum and 
the singular vectors of a POI ratings matrix
of customer votes for spots of France.
The fact that the matrix consists of 99.5\% missing values makes 
the spectrum highly dependent on how we handle the missing values.  
We computed the SVD of the full 0--1 ``implicit'' matrix of 
the visits without considering the rating.
For the ratings matrix, we used SGD, a popular approach that uses only 
the known values to compute the factors.
We observed that SGD and SVD factors are similar but SVD has stronger 
geo-localization. SVD singular vectors with highest eigenvalues 
are mostly correlated with a particular place.
As key practical observations, we found that imputing the missing ratings 
for the neighbors of visited places could increase the performance, and 
that defining fixed Geographic factors could improve SGD recommendation quality.

We expect that a broader analysis of a larger number
of similar type datasets of votes will allow to
gain better understanding of underlying
physical process and provide better recommendations for 
specific customers and spots.

\section{Acknowledgments}
We thank the representatives of NOMAO \cite{nomaocom}
and especially Estelle Delpech (NOMAO)
for providing us with the friendly access to the NOMAO datasets.
This research is supported in part by the EC FET Open project
``New tools and algorithms for directed network analysis''
(NADINE $No$ 288956).


\end{document}